\documentclass[%
 reprint,
superscriptaddress,
twocolumn,
nofootinbib,
 amsmath,amssymb,
 aps,
]{revtex4-2}

\usepackage{graphicx}
\usepackage{dcolumn}
\usepackage{bm}
\usepackage{breqn}
\usepackage{xcolor}
\usepackage{titlesec}
\usepackage[title]{appendix}

\begin{document}

\title{Fermi surface and Berry phase analysis for Dirac nodal line semimetals:\\ cautionary tale to SrGa$_2$ and BaGa$_2$}

\author{Yuxiang Gao}
\thanks{These authors contributed equally: Yuxiang Gao and Yichen Zhang}
\affiliation{Department of Physics and Astronomy and Smalley-Curl Institute$,$ Rice University$,$ Houston$,$ Texas 77005$,$ USA}

\author{Yichen Zhang}
\thanks{These authors contributed equally: Yuxiang Gao and Yichen Zhang}
\affiliation{Department of Physics and Astronomy and Smalley-Curl Institute$,$ Rice University$,$ Houston$,$ Texas 77005$,$ USA}
\author{Shiming Lei}
\thanks{current affiliation: Department of Physics$,$ Hong Kong University of Science and Technology$,$ Clear Water Bay$,$ Hong Kong$,$ China}
\affiliation{Department of Physics and Astronomy and Smalley-Curl Institute$,$ Rice University$,$ Houston$,$ Texas 77005$,$ USA}%
\author{Neil Harrison}
\affiliation{National High Magnetic Field Laboratory$,$ Los Alamos$,$ New Mexico 87545$,$ USA}
\author{Mun Keat Chan}
\affiliation{National High Magnetic Field Laboratory$,$ Los Alamos$,$ New Mexico 87545$,$ USA}
\author{Jonathan D. Denlinger}
\affiliation{Advanced Light Source$,$ Lawrence Berkeley National Laboratory$,$ Berkeley$,$ California 94720$,$ USA}
\author{Sergey Gorovikov}
\affiliation{%
Canadian Light Source Inc.$,$ University of Saskatchewan$,$ Saskatoon$,$ SK S7N 2V3$,$ Canada\\
}%
\author{Sanu Mishra}
\affiliation{Department of Physics and Astronomy and Smalley-Curl Institute$,$ Rice University$,$ Houston$,$ Texas 77005$,$ USA}
\author{Yan Sun}
\affiliation{Shenyang National Laboratory for Materials Science Institute of Metal Research$,$ Chinese Academy of Sciences}
\author{Ming Yi}
\affiliation{Department of Physics and Astronomy and Smalley-Curl Institute$,$ Rice University$,$ Houston$,$ Texas 77005$,$ USA}
\author{Emilia Morosan}
\email[corresponding author: E. Morosan ]{emorosan@rice.edu}
\affiliation{Department of Physics and Astronomy and Smalley-Curl Institute$,$ Rice University$,$ Houston$,$ Texas 77005$,$ USA}


\date{\today}

\begin{abstract}
A Berry phase of odd multiples of $\pi$ inferred from quantum oscillations (QOs) has often been treated as evidence for nontrivial reciprocal space topology. However, disentangling the Berry phase values from the Zeeman effect and the orbital magnetic moment is often challenging. In centrosymmetric compounds, the case is simpler as the orbital magnetic moment contribution is negligible. Although the Zeeman effect can be significant, it is usually overlooked in most studies of QOs in centrosymmetric compounds. Here, we present a detailed study on the non-magnetic centrosymmetric $\mathrm{SrGa_2}$ and $\mathrm{BaGa_2}$, which are predicted to be Dirac nodal line semimetals (DNLSs) based on density functional theory (DFT) calculations. Evidence of the nontrivial topology is found in magnetotransport measurements. The Fermi surface topology and band structure are carefully studied through a combination of angle-dependent QOs, angle-resolved photoemission spectroscopy (ARPES), and DFT calculations, where the nodal line is observed in the vicinity of the Fermi level. Strong de Haas-van Alphen fundamental oscillations associated with higher harmonics are observed in both compounds, which are well-fitted by the Lifshitz-Kosevich (LK) formula. However, even with the inclusion of higher harmonics in the fitting, we found that the Berry phases cannot be unambiguously determined when the Zeeman effect is included. We revisit the LK formula and analyze the phenomena and outcomes that were associated with the Zeeman effect in previous studies. Our experimental results confirm that $\mathrm{SrGa_2}$ and $\mathrm{BaGa_2}$ are Dirac nodal line semimetals. Additionally, we highlight the often overlooked role of spin-damping terms in Berry phase analysis. 

\end{abstract}

\maketitle

\section{Introduction}
Topological semimetals are gapless solid state phases with band touchings near the Fermi level, harboring nontrivial topological invariants, often with linear low energy dispersions \cite{Hasan2010, Burkov2017, Armitage2018, Bernevig2018, Hu2019}. These materials can be further classified according to the dimension and degeneracies of the band crossings~\cite{Burkov2017, Armitage2018, Bernevig2018, Hu2019, Bradlyn2016, Fang2016, Chang2018, Xie2021}. Dirac nodal line semimetals (DNLSs), which display four-fold degenerate nodal lines, have recently garnered significant interest~\cite{Hu2019, Fang2016}, since they are predicted to be the parent phase of many topological phases. Breaking specific symmetries or turning on spin-orbit coupling (SOC) provides routes to tune Dirac nodal lines into Weyl nodal lines, Dirac points, Weyl points, and Kramers nodal lines~\cite{Hu2019, Fang2016, Lei2022,Zhang2023,Zhang2025}. The topologically nontrivial Dirac nodal lines bring about various exotic magnetotransport and spectroscopic signatures~\cite{Schoop2016, Hu2016, Ali2016, Singha2017, 2Singha2017, Schoop2018, Chiu2019}, such as large non-saturating magnetoresistance and drumhead surface states observed in ARPES.   

The nontrivial $\pi$ Berry phase in QOs has been esteemed as smoking gun evidence for nontrivial reciprocal space topology~\cite{Hu2019, Hu2016, Singha2017, He2014, Fei2017, Huang2017, Wang2018, Ma2021, Cheng2021}. Such a Berry phase originates from the band-touching point (line) and is accumulated in the cyclotron motion of Dirac or Weyl fermions if the cyclotron orbit encloses the band crossing. The nontrivial $\pi$ Berry phase seems to be the most prevalent transport evidence compared to other phenomena: the anomalous Hall effect (AHE) is not observed in most non-magnetic topological materials, large non-saturating magnetoresistance and chiral anomaly might have other origins (electron-hole compensation and current jetting effect, respectively)~\cite{Tian2018, Zeng2016, Guo2018, Arnold2016, LiangS2018}. It is important to note that the Berry phase in three-dimensional topological materials may deviate from the exact $\pi$ value due to the following reasons: (1) detailed Fermi surface geometry and magnetic field orientation \cite{He2014,Li2018,YangY2021}; (2) deviation from perfect linear dispersion \cite{Taskin2011}; (3) a gap opening at the band crossing \cite{Mikitik1999,LuH2010,Wright2013}.

However, the Berry phase extracted by QO analysis may consist of additional contributions and should instead be written as a phase shift $\lambda$, with the Berry phase being one of the contributions. In recent studies, the analysis has been performed mainly by fitting the simplified LK formula or using a Landau-level (LL) fan plot \cite{Shoenberg1984,Kwan2020}. The former method fits the QOs while discarding the higher harmonics and the spin-damping term. The latter method assigns half-integer/integer values (LL indices n) to the peaks/valleys of the fundamental oscillations. The Berry phase is reflected in the intercept of the linear plot of the inverse external magnetic field 1/$\mu_0H$ \textit{vs} the LL index n, which also neglects higher harmonics and the spin-damping term. In both methods, one only accurately determines the phase shift $\lambda$ and \textit{not} the Berry phase, with $\lambda~=~\phi_B~+~\phi_R~+~\phi_Z$, where $\phi_B$, $\phi_R$, and $\phi_Z$ are Berry phase, orbital magnetism and Zeeman coupling terms~\cite{Alexandradinata2018}. Therefore, a measured $\pi$ phase shift does not necessarily represent the Berry phase, and further analysis or probes to tackle each contribution separately are required to unambiguously determine the Berry phase. In recognition of this need in the Berry phase analysis, the inclusion of higher harmonics in the QOs has been proposed as a way to separate different contributions, which paves a more rigorous approach \cite{Alexandradinata2018}. In addition, de Haas-van Alphen (dHvA) oscillations are preferred to Shubnikov-de Haas (SdH) oscillations: magnetization is a thermodynamic variable given by an exact expression known as the LK formula \cite{Shoenberg1984}. Conversely, because resistivity or conductivity is not a thermodynamic variable, the SdH oscillations are often complicated by different scattering mechanisms~\cite{Shoenberg1984, Liu2017}. Therefore, a topological semimetal that manifests strong dHvA oscillations with higher harmonics is paramount to establishing a careful study of the Berry phase. 

In this work, we present magnetotransport, angle- and temperature-dependent dHvA oscillations, and ARPES results on both $\mathrm{SrGa_2}$ and $\mathrm{BaGa_2}$. The electrical transport measurements confirm the good crystal quality of both materials and provide evidence of nontrivial topology. We comprehensively establish the Fermi surface of both compounds using angle-dependent dHvA and ARPES measurements, as well as DFT calculations. Via careful analysis of the temperature-dependent QOs with the inclusion of higher harmonics and Zeeman effect through LK fitting, we find that it is not possible to unambiguously determine the Berry phase and Landé g-factor through QOs alone. Although previous works showed ways to estimate Landé g-factor through peak splitting in QOs \cite{Nimori1994, Narayanan2015, Cao2015, HuJ2016, Liu2016, Martin2018, Xiang2021}, our theoretical deduction suggests that those peak splittings might have different origins. Nonetheless, we demonstrate that $\mathrm{SrGa_2}$ and $\mathrm{BaGa_2}$ are Dirac nodal line semimetal materials through a combination of experimental and theoretical techniques and highlight the importance of the Zeeman effect in the Berry phase analysis through QOs which makes the precise estimation of the Berry phase challenging. 

\section{Methods}

The single crystals of $\mathrm{SrGa_2}$ and $\mathrm{BaGa_2}$ were grown by the flux method. The starting elements with compositions Sr:Ga = 50:50/Ba:Ga = 40:60 were placed in tantalum crucibles and sealed in evacuated quartz ampoules. Then the ampoules were heated between 1050 and 950\textdegree C, held at the maximum temperature for 4 hours, before cooling down to temperatures between 750 and 610\textdegree C at $\sim$ 2\textdegree C/h. The $\mathrm{SrGa_2}$ and $\mathrm{BaGa_2}$ crystals were obtained after decanting excess flux in a centrifuge. The atomic composition was checked by energy-dispersive X-ray spectroscopy (EDS). The crystal structures were determined by powder X-ray diffraction in a Bruker D8 Advance X-ray diffractometer with Cu K-$\alpha$ radiation. Rietveld refinements were done using TOPAS software. 

The electrical transport measurements were conducted in a Quantum Design (QD) Dynacool PPMS-14~T system using a four-contact method. The magnetization up to 9 T was measured in a QD Dynacool PPMS using the VSM option. High-field magnetic torque measurements were measured in a 65~T pulsed magnet at the National High Magnetic Field Laboratory, Los Alamos. 

ARPES measurements for $\mathrm{SrGa_2}$ were taken at the Beamline 4.0.3 (MERLIN) of the Advanced Light Source, equipped with a SCIENTA R8000 analyzer with vertical slit relative to the ground. The temperature and vacuum during measurements were kept at 22 K and below 5$\times$10$^{-11}$ Torr. Photon energies between 30 and 120 eV in linear horizontal polarization were scanned on the in-situ cleaved $\mathrm{SrGa_2}$. ARPES experiments on $\mathrm{BaGa_2}$ were conducted at the Quantum Materials Spectroscopy Centre beamline of the Canadian Light Source, equipped with a R4000 electron analyzer, where the slit of the analyzer is horizontal relative to the ground. The BaGa$_2$ samples were cleaved in-situ and measured under a vacuum condition better than 6$\times$10$^{-11}$ Torr and temperature stabilized at 15 K. Photon energy scan on $\mathrm{BaGa_2}$ covers a range from 56 to 120 eV using linear vertical polarization perpendicular to the analyzer slit direction. Energy and angular resolution for all the measurements were set to be better than 25 meV and 0.1$^\circ$, respectively.

\begin{figure*}
\includegraphics[width=0.95\textwidth]{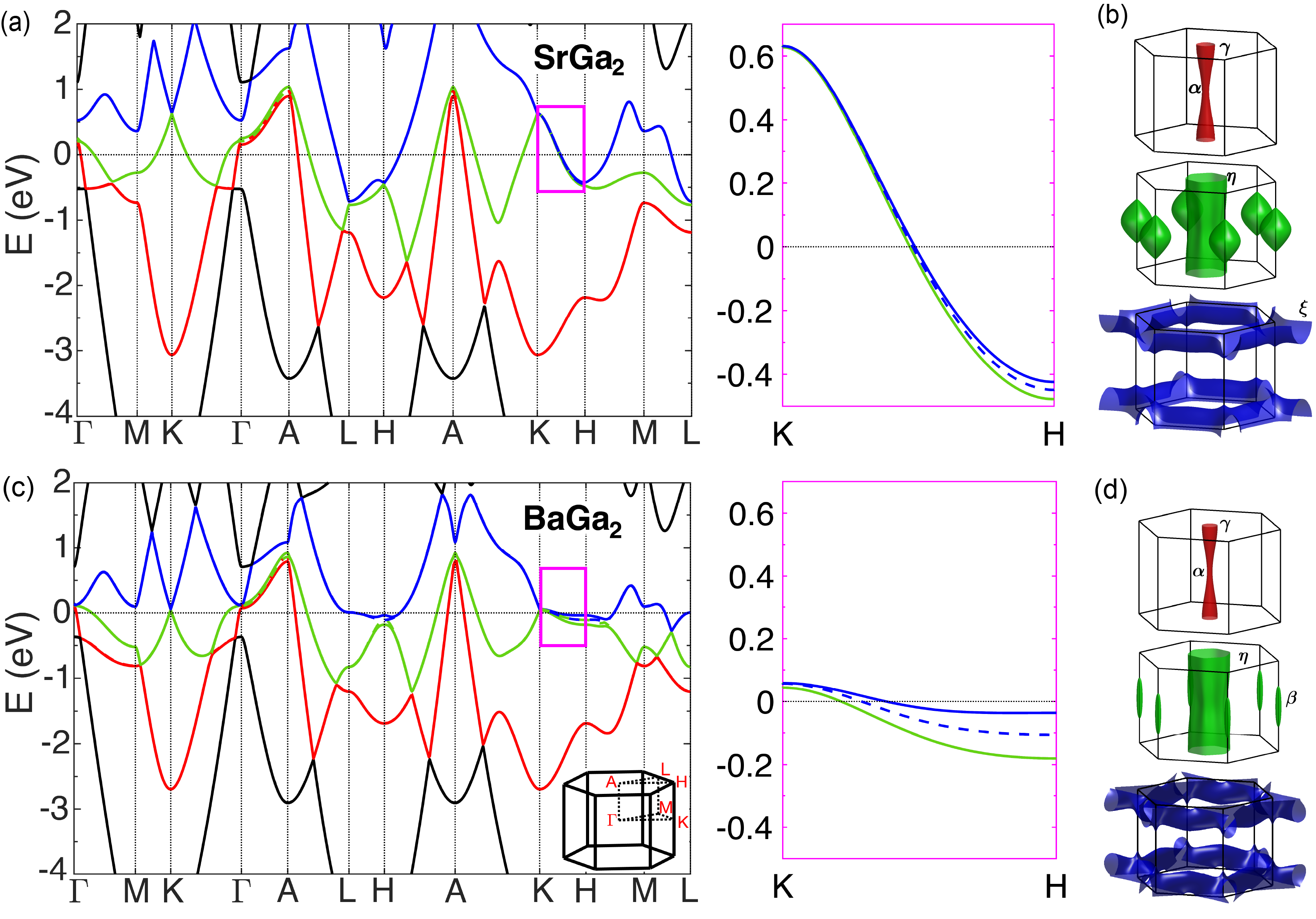}
\caption{\label{fig:4} Calculated band structures and Fermi surfaces of $\mathrm{SrGa_2}$ and $\mathrm{BaGa_2}$. (a,c) Left: Band structure calculations with SOC (solid lines) and without SOC (dashed lines) of $\mathrm{SrGa_2}$ (a) and $\mathrm{BaGa_2}$ (c) along high-symmetry paths. The bands that cross the Fermi energy are highlighted in different colors. The inset of (c) shows the Brillouin zone with labels of high-symmetry points. The magenta boxes highlight the Dirac nodal lines along the K-H direction. Right: Zoomed-in view of the Dirac nodal lines along the K-H direction. A small gap develops between the two bands that form the Dirac nodal lines because of SOC. (b,d) Calculated Fermi surfaces for $\mathrm{SrGa_2}$ (b) and $\mathrm{BaGa_2}$ (d) with SOC included. The colors of the Fermi pockets are consistent with those in (a,c). The labels represent the QO frequencies as shown in Fig.~\ref{fig:3} and Fig.~\ref{fig:5}.
}
\end{figure*} 

The electronic band structures of $\mathrm{SrGa_2}$ and $\mathrm{BaGa_2}$ were calculated based on DFT using the code of Vienna ab-initio simulation package (VASP) \cite{vasp} with a projected augmented wave potential. The exchange and correlation energies were considered at the level of generalized gradient approximation (GGA), following the Perdew-Burke-Ernzerhof parametrization scheme \cite{pbe}. All the calculations were performed based on experimental lattice constants. 

The temperature-dependent dHvA oscillations were fitted using the LK formula with the inclusion of higher harmonics~\cite{Shoenberg1984}:
\begin{multline}
\label{eq1}
\Delta M \propto -B^\gamma \sum_p \sum_{r=1}^{\infty}\frac{1}{r^{3/2}}R_T R_D R_S \\
\sin\Bigg(2\pi \bigg[r\bigg(\frac{F}{B}-\gamma+\beta\bigg)+\delta\bigg]\Bigg)
\end{multline}
where $R_T={raT \mu}/{sinh(raT \mu)}, R_D=\exp(-{raT_D\mu}/{B})$ and $R_S=\cos({r\pi g\mu}/{2})$ terms are the temperature, field, and spin damping terms. $p$ and $r$ are the pocket and harmonic indices, $\mu={m_{eff}}/{m_0}$ is the ratio of the effective mass $m_{eff}$ to the free electron mass $m_0$, $a=(2{\pi}^2 k_B m_0)/ (\hbar e)\approx14.69$ T/K, $T_D$ is the Dingle temperature and $F$ is the frequency. $\gamma=0, \delta=0$ for a 2D pocket, and $\gamma=1/2, \delta=\pm 1/8$ for a minimum/maximum cross-section of a 3D electron/hole pocket, respectively. $\beta=\phi_B/2\pi$ and $\phi_B$ is the Berry phase \cite{Hu2019}.

\begin{figure}
\includegraphics[width=0.45\textwidth]{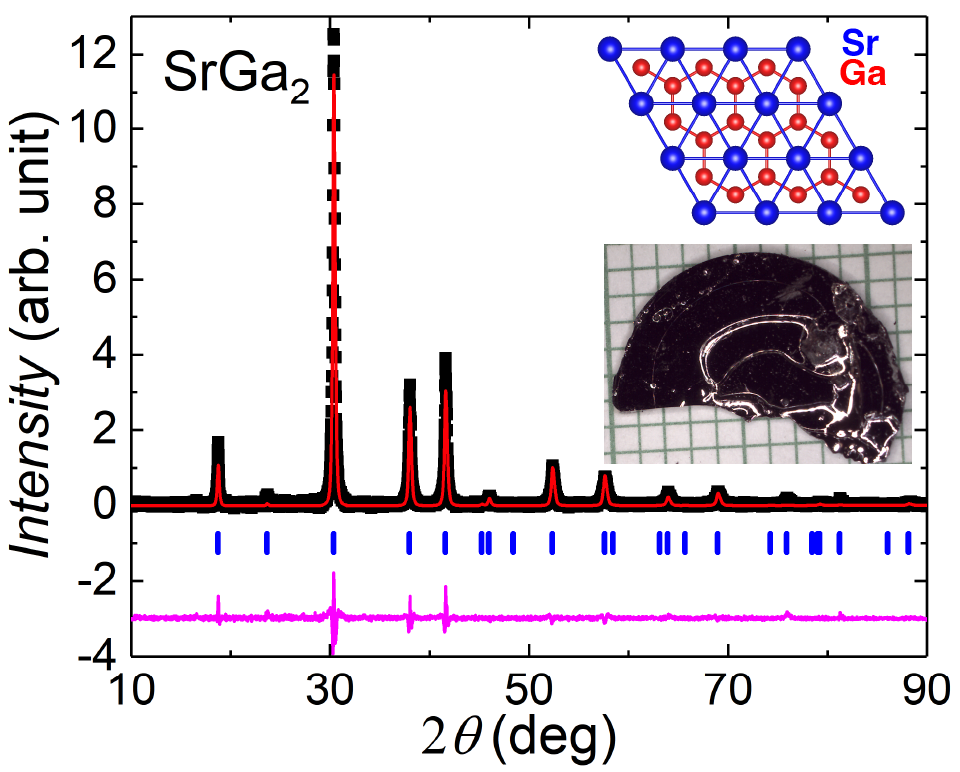} 
\caption{\label{fig:1} Powder X-ray diffraction pattern of $\mathrm{SrGa_2}$. The upper inset is the crystal structure projection along the c-axis showing honeycomb layers of Ga atoms and triangular layers of alkalines. The lower inset shows one piece of single crystal, with each grid equal to 1mm.}
\end{figure}
\begin{figure*}
\includegraphics[width=0.96\textwidth]{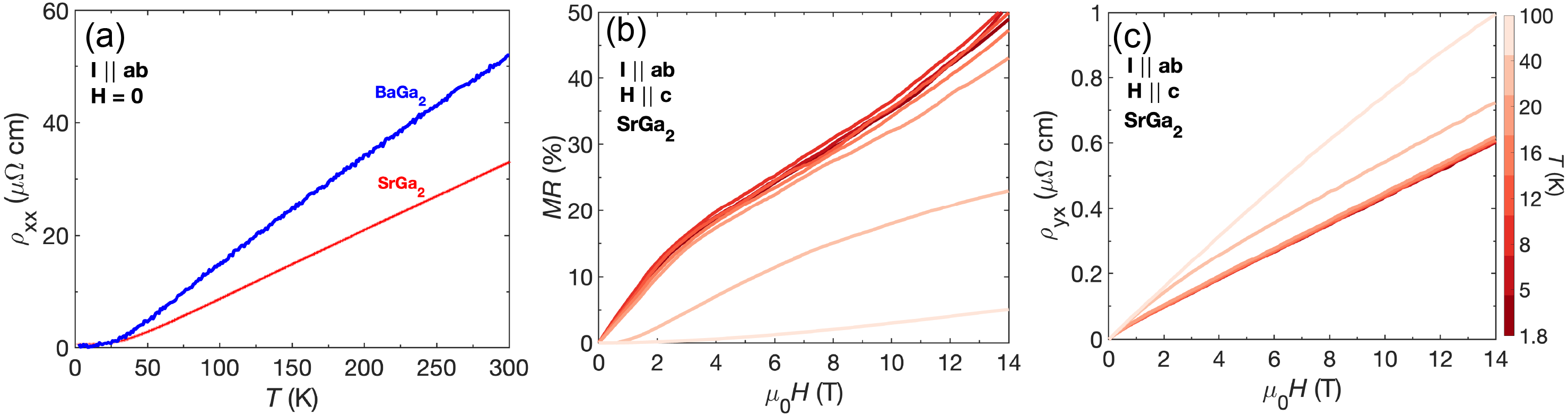}
\caption{\label{fig:2} In-plane resistivity and Hall resistivity of $\mathrm{SrGa_2}$ and $\mathrm{BaGa_2}$.  (a) In-plane zero field resistivity $\rho_{xx}$ of $\mathrm{SrGa_2}$ and $\mathrm{BaGa_2}$. (b,c) In-plane MR and Hall resistivity of $\mathrm{SrGa_2}$ at different temperatures .}
\end{figure*}

\section{Results}
\subsection{DFT calculations}
We identify $\mathrm{SrGa_2}$ and $\mathrm{BaGa_2}$ as good material platforms for this study based on electronic band structure calculations. The electronic structures along the high symmetry paths, with or without the SOC, are shown in Fig. \ref{fig:4}(a,c). The band dispersions of $\mathrm{SrGa_2}$ and $\mathrm{BaGa_2}$ are similar. In both compounds, three bands (depicted in red, green, and blue colors) cross the Fermi energy and form four Fermi pockets, as shown in Fig. \ref{fig:4}(b,d): $\alpha/\gamma$ (red), $\beta$(green), $\eta$(green), $\xi$(blue). According to DFT calculations, the only topological band crossing the Fermi energy is a Dirac nodal line along K-H, highlighted in the magenta box in Fig. \ref{fig:4}(a,c). These Dirac nodal lines will open a small gap when SOC is included. The gap is much smaller in SrGa$_2$ than in BaGa$_2$, due to the stronger SOC of Ba. Although the gap in BaGa$_2$ is relatively large along K-H, the gap at the K point is small ($\sim$10 meV), and the low-energy excitation can still be described by the Dirac equation and possesses the characteristic of Dirac fermions \cite{Xu2020}. Therefore, it is more suitable to classify BaGa$_2$ as a Dirac semimetal. Nonetheless, only the QOs from the $\beta$ pocket, elongated along the K-H direction (green in Fig. \ref{fig:4}(d)), are expected to manifest a nontrivial $\pi$ Berry phase, while the QOs from other pockets should yield a trivial zero Berry phase.

\subsection{Structural characterization}
$\mathrm{SrGa_2}$ and $\mathrm{BaGa_2}$ crystallize in the {\it P6/mmm} space group ($\mathrm{AlB_2}$ type), which consists of Ga honeycomb layers sandwiched between triangle layers of alkaline atoms (Fig.~\ref{fig:1}(a)). The lattice parameters were determined from the Rietveld refinements on powder X-ray diffraction, yielding a = 4.341~{\AA}, c = 4.734~{\AA} for $\mathrm{SrGa_2}$ and a = 4.440~{\AA}, c = 5.082~{\AA} for $\mathrm{BaGa_2}$. The refined lattice parameters imply that the c/a ratio of $\mathrm{BaGa_2}$ is 5$\%$ larger than that of $\mathrm{SrGa_2}$, which suggests that $\mathrm{BaGa_2}$ is more two-dimensional than $\mathrm{SrGa_2}$.

\subsection{Magnetotransport measurements}
Previously, magnetotransport measurements \cite{Xu2020} showed evidence of nontrivial topology in $\mathrm{BaGa_2}$, where a large non-saturating transverse magnetoresistance (MR), negative longitudinal MR (interpreted as inter-layer quantum tunneling), and nonzero Berry phases in QOs were observed \cite{Xu2020}. The hole carrier nature, as indicated by Hall measurements, ruled out an electron-hole compensation being responsible for the large non-saturating MR \cite{Xu2020}.

Similar behavior is found in $\mathrm{SrGa_2}$. The in-plane resistivity shows a metallic behavior, with residual resistivity ratio RRR~=~$\rho$(300K)/$\rho$(1.8K)~=~55 as shown in Fig. \ref{fig:2}(a). The Hall resistivity $\rho_{yx}$ shown in Fig. \ref{fig:2}(c) suggests that the dominant carrier is hole-like, which can be further fit by a two-band model with two hole carriers of different mobilities (Fig.~\ref{fig:SI1}(b,c,d)). The MR of SrGa$_2$ shown in Fig. \ref{fig:2}(b) is non-saturating and reaches 50\% at 14~T and 1.8~K, with no signature of SdH oscillations. The behaviors of MR and Hall in $\mathrm{SrGa_2}$ are therefore similar to $\mathrm{BaGa_2}$ ~\cite{Xu2020}. Planar Hall Effect (PHE) is observed in $\mathrm{SrGa_2}$, indicated by the sin(2$\psi$) behavior in Fig.~\ref{fig:SI1}(a). The PHE, large non-saturating MR, and non-compensated carrier nature indicate that $\mathrm{SrGa_2}$ likely hosts nontrivial topology, similar to $\mathrm{BaGa_2}$  \cite{Hu2019,Xu2020}.

\begin{figure*}
\includegraphics[width=0.96\textwidth]{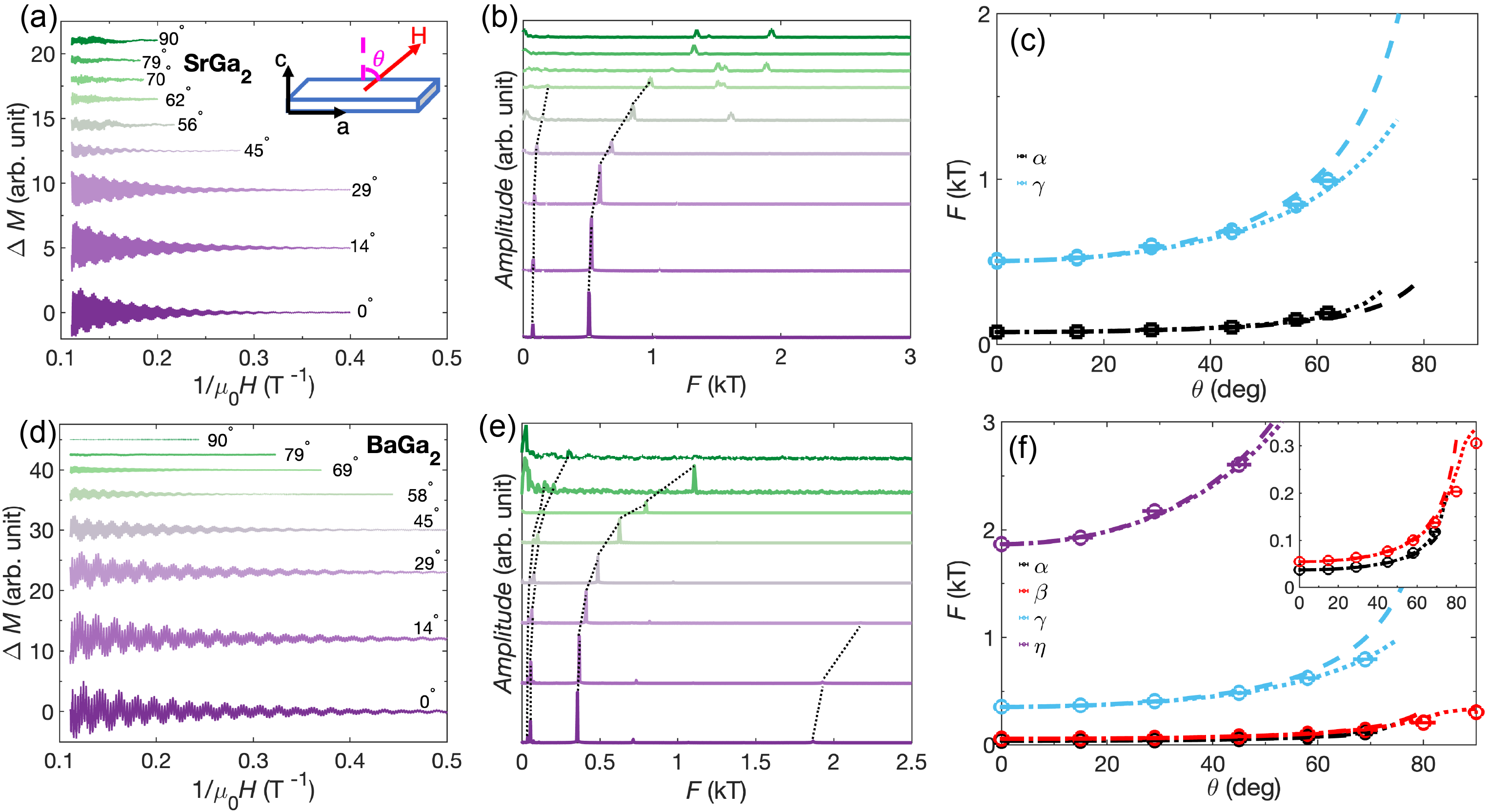}
\caption{\label{fig:3} Angle-dependent dHvA oscillations of $\mathrm{SrGa_2}$ and $\mathrm{BaGa_2}$. (a,d) Angle-dependent dHvA oscillations of $\mathrm{SrGa_2}$ (a) and $\mathrm{BaGa_2}$ (d). The inset of (a) is a sketch of the measurement. (b,e) FFT spectra of $\Delta M$ of $\mathrm{SrGa_2}$(b) and $\mathrm{BaGa_2}$(e). The FFT spectra at $\theta$ = 79\textdegree and 90\textdegree for $\mathrm{BaGa_2}$ are amplified by 40 and 60 times, respectively. An offset to the FFT amplitude has been applied for better visualization. Dashed lines are guides to the eye for the frequency changes. (c,f) QO frequencies from fundamental oscillations as a function of angle for $\mathrm{SrGa_2}$ (c) and $\mathrm{BaGa_2}$ (f). Symbols are extracted values from FFTs in (b,e), dotted lines are values from DFT, and dashed lines are estimated values of a cylindrical Fermi surface. }
\end{figure*}

\subsection{Angle-dependent dHvA oscillations}

Angle-dependent quantum oscillations were used to determine the Fermi surface and its underlying topology. The Onsager relationship illustrates the connection between the QO frequency F and the extremal Fermi surface cross-section $S_{ext}$:  $F={\hbar S_{ext}}/{2\pi e}$, where $\hbar$ is the reduced Planck constant and $S_{ext}$ is the extremal Fermi surface cross-section perpendicular to the magnetic field. We determine F from dHvA oscillations. The experimental geometry is shown in the inset of Fig.~\ref{fig:3}(a). Strong dHvA oscillations can be found in the isothermal magnetization of $\mathrm{SrGa_2}$ and $\mathrm{BaGa_2}$ after a smooth background subtraction, as shown in Fig.~\ref{fig:3}(a,d), and they persist for all magnetic field orientations from  $\theta$ = 0\textdegree~(out-of-plane magnetic field) to $\theta$ = 90\textdegree~(in-plane magnetic field). The oscillation amplitude decreases monotonically in $\mathrm{BaGa_2}$ from $\theta$ = 0\textdegree~ to  90\textdegree, while in $\mathrm{SrGa_2}$ it decreases from $\theta$ = 0\textdegree~ to 45\textdegree~ and is nearly constant in the range of $\theta$~=~45\textdegree~to 90\textdegree. Such angular dependence of the QO amplitude underscores the picture of $\mathrm{BaGa_2}$ being more two-dimensional than $\mathrm{SrGa_2}$. Such a difference in the electronic dimensionality can be inferred from the band dispersion along the k$_z$ direction, which is best illustrated in the K-H direction, as shown in the right panel of Fig. 1(a,c).

The QO frequencies at different field orientations are obtained from the fast Fourier transform (FFT) in Fig.~\ref{fig:3}(a,d), and are shown in Fig.~\ref{fig:3}(b,e). These frequencies are further compared with the values of cylindrical Fermi pockets to estimate the shape of the corresponding Fermi pockets in QOs. For a cylindrical Fermi surface elongated along the c axis and field-oriented at an angle $\theta$ away from the c axis, the cross-sectional area is ${S}/{\cos(\theta)}$, where S is the basal area of the cylinder. The comparison can be found in Fig.~\ref{fig:3}(c,f). We found that $\alpha$ in $\mathrm{SrGa_2}$ and $\mathrm{BaGa_2}$ corresponds to necks [$F(\theta)>F(0)/\cos(\theta)$], while the frequencies $\gamma$ in $\mathrm{SrGa_2}$, and $\beta$ and $\gamma$ in $\mathrm{BaGa_2}$ are bellies [$F(\theta)<F(0)/\cos(\theta)$], as demonstrated in Fig.~\ref{fig:3}(c,f). 

We further compared the frequencies from the FFT spectra with cross-sections from DFT calculations. Based on the shape and size of the Fermi pockets as in Fig.~\ref{fig:3}(b,d), we linked the extremal cross-sections of the Fermi pockets and the QO frequencies in the following way: in $\mathrm{SrGa_2}$, the $\alpha$ and $\gamma$ frequencies are the minimum and maximum cross-sections of the Fermi pocket shown in red in Fig. \ref{fig:4}(b); in $\mathrm{BaGa_2}$, the $\alpha$ and $\gamma$ frequencies are the minimum and maximum cross-sections of the Fermi pocket shown in red, the $\beta$ frequency is the spindle-like Fermi pocket at K point shown in green, the $\eta$ frequency is the maximum cross-section of the Fermi pocket along the $\Gamma -A$ direction shown in green in Fig. \ref{fig:4}(d). The correspondence is further supported by a perfect match between cross-sections estimated from calculations and frequencies from experimental results as shown in Fig.~\ref{fig:3}(c,f). Furthermore, under the same measurement geometry, we performed magnetic torque measurements under pulsed magnetic fields up to 65 T for both $\mathrm{SrGa_2}$ and $\mathrm{BaGa_2}$. The same correspondence is established (Fig. \ref{fig:SI4}). Lastly, we trace the $\eta$ frequency in $\mathrm{SrGa_2}$ back to the maximum cross-section of the Fermi pocket along the $\Gamma -A$ direction shown in green ($\eta$ in Fig. \ref{fig:4}(b)). In the previous work from \cite{Xu2020}, a different correspondence is built on $\mathrm{BaGa_2}$, albeit without direct comparison between calculations and experiments \cite{Xu2020}.

In particular, the Fermi pocket $\xi$ shown in blue in Fig. \ref{fig:4}(d) is substantively different from other pockets in red and green: it spans the entire $k_x-k_y$ plane, while the rest of the pockets are elongated along the c-axis. According to DFT calculations, if the field is close to the c direction, the QOs from this pocket are forbidden, since the cross section perpendicular to the magnetic field is open. Instead, if an in-plane magnetic field is applied, the QO frequencies that originate from this pocket are sensitive to the field orientation and should be observed in corresponding measurements. Based on the understanding of the Fermi surface topology from DFT and SOC, only the $\beta$ pocket in $\mathrm{BaGa_2}$ is related to the Dirac nodal line and has nontrivial topology.

We also measured dHvA oscillations with an in-plane magnetic field on $\mathrm{SrGa_2}$. The results can be found in Fig.~\ref{fig:SI2}, with the measurement geometry shown in (a). The shape of the background-subtracted results (Fig.~\ref{fig:SI2}(a)) and its FFT spectrum (Fig.~\ref{fig:SI2}(b)) shows an orientation dependence, as expected. Furthermore, a good match is shown in Fig.~\ref{fig:SI2}(c), confirming the observation of QOs from the blue pocket in $\mathrm{SrGa_2}$. Note that in $\mathrm{SrGa_2}$, QOs of similar frequencies are observed at $\theta > $45\textdegree. We suspect that those oscillations originate from the Fermi pocket $\xi$.

\subsection{ARPES measurements}
\begin{figure*}
\includegraphics[width=0.9\textwidth]{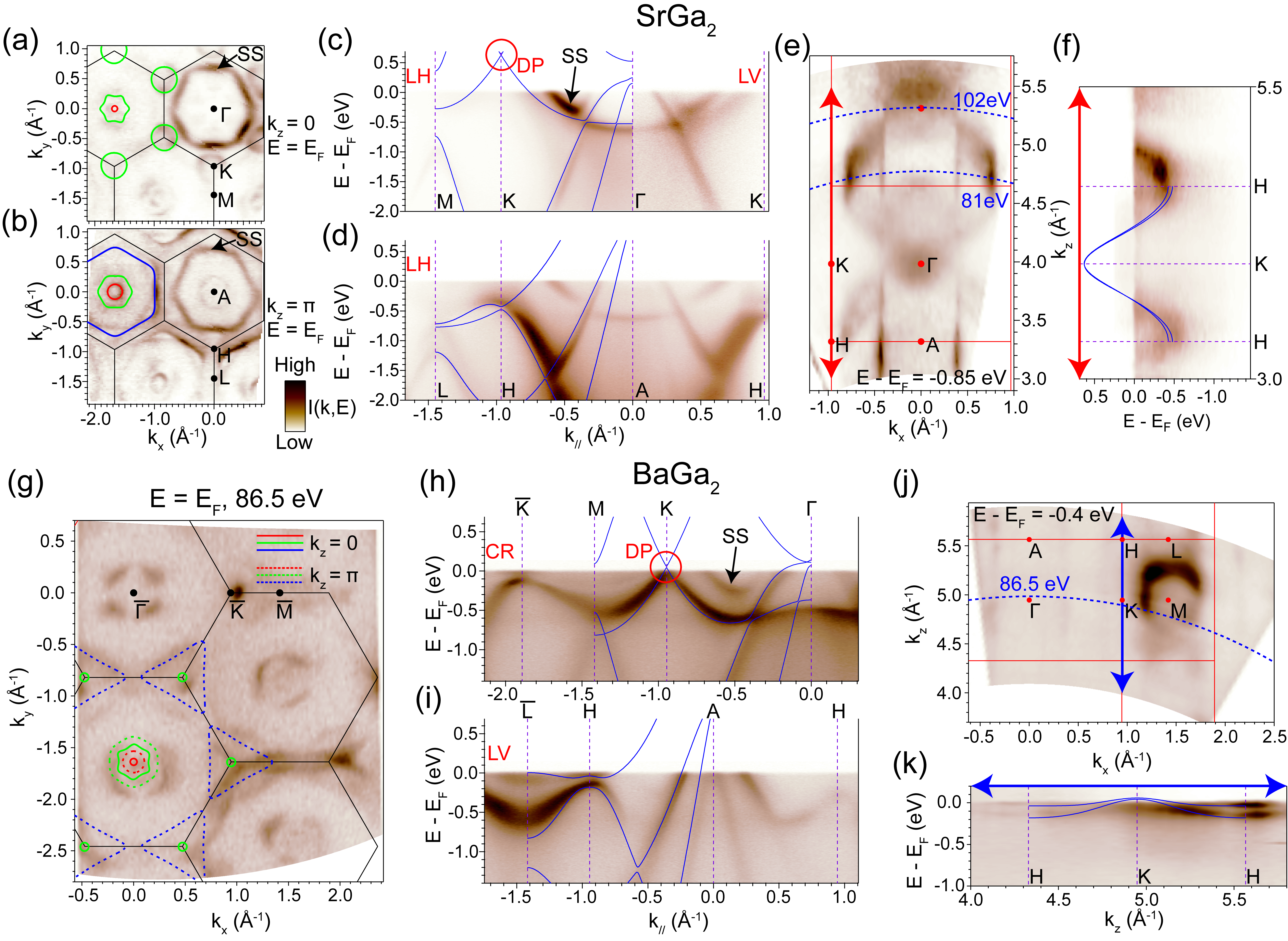}
\caption{\label{fig:arpes} ARPES measurements of the electronic structure of (a)-(f) SrGa$_2$ and (g)-(k) BaGa$_2$. (a)-(b) Fermi surface of SrGa$_2$ near $k_z$ = 0 (102 eV) and $\pi$ (81 eV) measured by linear horizontal (LH) photons. Fermi pockets calculated from density functional theory (DFT) are overlaid on top, the colors of which correspond to those assigned in Fig. \ref{fig:4}(b). (c)-(d) High symmetry band dispersions along $\Gamma$-$K$-$M$ and $A$-$H$-$L$ with DFT bands superimposed for comparison. The Dirac point (DP) at $K$ is circled in red. (e) Constant energy contour at $E-E_{\rm F}=-0.85$ eV in the $k_z-k_x$ plane through varying the photon energy, where the photon energies used in (a) and (b) are marked by the dashed blue curves. (f) Band dispersions along the out-of-plane $H$-$K$-$H$ direction showing spin-orbit-coupling(SOC)-split nodal lines of SrGa$_2$. The cut direction is marked by the vertical red arrow in (e). (g) Fermi surface of BaGa$_2$ measured with 86.5 eV LH polarized photons showing both $k_z$ = 0 (solid contours) and $k_z$ = $\pi$ pockets (dashed contours) predicted by DFT. (h)-(i) Equivalent cuts as in (c) and (d) but for BaGa$_2$. The $\overline{K}$ and $\overline{L}$ notation in (h) and (i) is due to the large curvature of the cut shown by the dashed curve in (j), so the second ``$K$" or ``$L$" deviates from the high-symmetry plane significantly. (j) $k_x$-$k_z$ constant energy contour of BaGa$_2$ at $E-E_{\rm F}=-0.4$ eV. From the blue double arrow along $H$-$K$-$H$, the band dispersions containing the SOC-split Dirac nodal line are extracted and shown in (k). SS: surface states. LH: linear horizontal. LV: linear vertical. CR: circular right. These light polarizations are denoted for different momentum segments of the displayed band dispersions in (c), (d), (h), and (i) to showcase a more complete band structure.}  
\end{figure*}

The electronic band structure can be directly visualized by ARPES measurements. In Fig.~\ref{fig:arpes}, we present a direct comparison of electronic band dispersions between SrGa$_2$ and BaGa$_2$. First, the measured Fermi surfaces at $k_z$ = 0 and $\pi$ are shown in Fig.~\ref{fig:arpes}(a,b) for SrGa$_2$, and Fig.~\ref{fig:arpes}(g) for BaGa$_2$, overlaid with their corresponding DFT calculations following the color code defined in Fig.~\ref{fig:4}. The observed Fermi surfaces account for all the pockets that cross the Fermi energy $E_{\rm F}$ predicted by DFT. In addition, a hexagonal Fermi pocket, persistent at both $k_z$ = 0 and $\pi$ and absent in bulk DFT calculations, can be observed in SrGa$_2$, which we identify as a surface state (SS). The measured in-plane band dispersions along high-symmetry directions are shown in Fig.~\ref{fig:arpes}(c,d) for SrGa$_2$ and in Fig.~\ref{fig:arpes}(h,i) for BaGa$_2$, where polarization-dependent ARPES has been employed and organized together to reveal the electronic bands more thoroughly. The excellent consistency between the overlaid bulk DFT calculations and ARPES spectra allows us to identify the Dirac points (DPs) in SrGa$_2$ and BaGa$_2$. In particular, the DP in BaGa$_2$ located at $K$ is very close to the Fermi level. 

To directly visualize the Dirac nodal lines, we carried out photon energy-dependent ARPES measurements to probe the out-of-plane momentum direction. Figure~\ref{fig:arpes}(e,j) shows the $k_z-k_x$ maps of SrGa$_2$ and BaGa$_2$, respectively. The clear out-of-plane periodicity allows us to pin down the high-symmetry $k_z$ planes. From this we can extract the Dirac nodal lines of SrGa$_2$ and BaGa$_2$ along the out-of-plane $K$-$H$ direction (Fig.~\ref{fig:arpes}(f,k)). The nodal line dispersions show excellent agreement with the overlaid bulk DFT calculations for both compounds. In BaGa$_2$, due to the stronger SOC, the Dirac nodal line clearly shows a larger splitting away from the K point. In particular, the bandwidth of the nodal lines differs significantly between the two compounds, with a much smaller bandwidth in BaGa$_2$, consistent with the larger c/a ratio of BaGa$_2$ and hence a stronger two-dimensionality compared to the Sr analogue, resulting in flat Dirac nodal lines. 
The electronic band dispersions directly observed by ARPES provide a complete view of the electronic dispersions in SrGa$_2$ and BaGa$_2$ from the three dimensions $k_x$, $k_y$, and $k_z$. This helps in understanding the Fermi surface topology and offers solid support to the results from angle-dependent QO measurements and DFT calculations.

\subsection{Temperature dependent dHvA oscillations} 
\begin{figure*}
\includegraphics[width=0.96\textwidth]{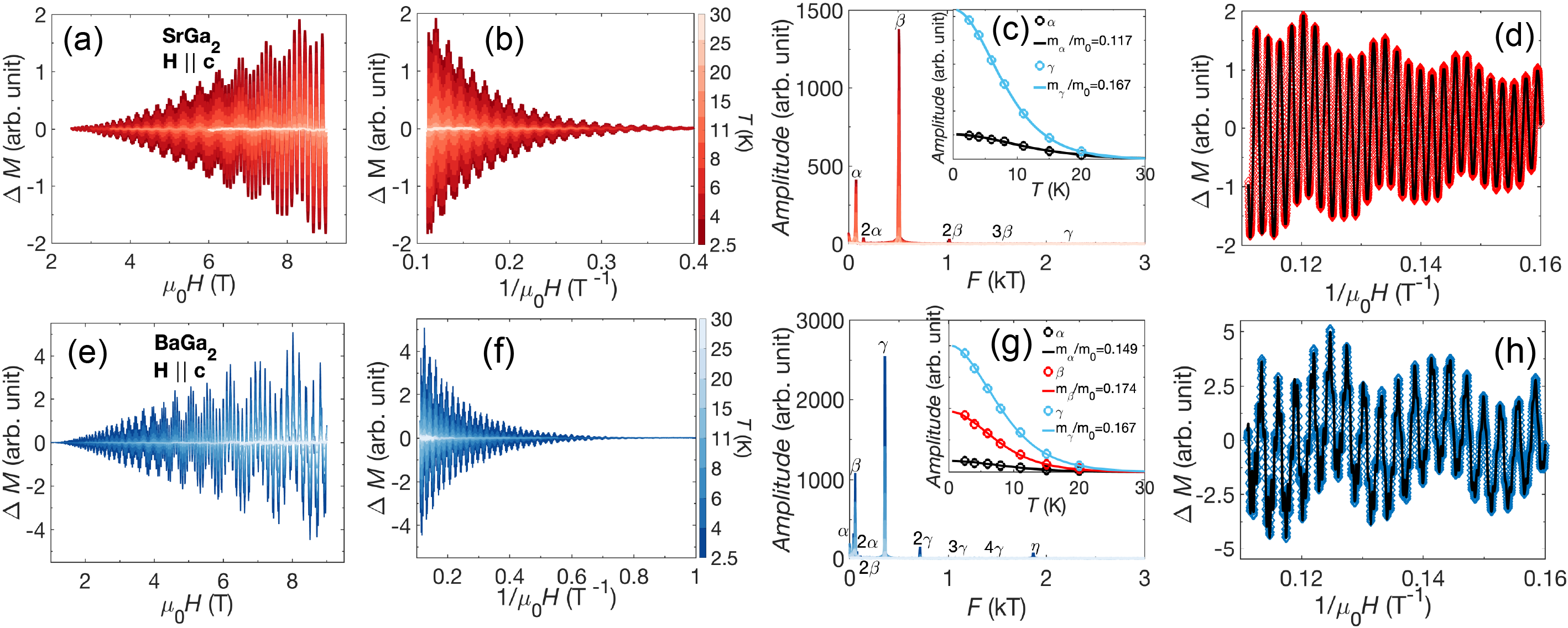} 
\caption{\label{fig:5} Temperature-dependent dHvA oscillations of $\mathrm{SrGa_2}$ and $\mathrm{BaGa_2}$. (a,b,e,f) The oscillatory component $\Delta M$ after background subtraction as a function of $\mu_0H$ (a,e) and 1/$\mu_0H$ (b,f). (c,g) FFT spectra of $\Delta M$ as in (b,f). The effective mass is determined by fitting the thermal factor of the LK formula to the oscillation amplitude (inset). (d,h) $\Delta M$ (red/blue symbols) to the LK fitting (black lines) at 2.5 K.}
\end{figure*}

With a detailed understanding of the Fermi surface, we investigated QOs at different temperatures to analyze the Berry phase of each Fermi pocket. The field is applied along the c axis (the out-of-plane direction). In sharp contrast to the resistivity up to 14~T at 1.8~K where QOs are absent, dHvA oscillations are observed up to 30~K and down to 2~T/1~T in $\mathrm{SrGa_2}$/$\mathrm{BaGa_2}$ [Fig.~\ref{fig:5}(a,e)], respectively. Three fundamental frequencies in $\mathrm{SrGa_2}$ are identified in the FFT spectra of the background-subtracted signal. The fundamental oscillation frequencies of $\mathrm{BaGa_2}$ are very close to the values reported in~\cite{Xu2020}. In addition to the fundamental frequencies, higher harmonics with considerable intensities can be distinguished from the background, up to the third harmonic in $\mathrm{SrGa_2}$ and the fourth harmonic in $\mathrm{BaGa_2}$, as indicated in Fig.~\ref{fig:5}(c,g). 

\begin{table*}

    \centering
    \begin{ruledtabular}
    \begin{tabular}{cccccccccc}
    &\multicolumn{9}{c}{$\mathrm{SrGa_2}$}\\
     Index&$F(T)$&$T_D(K)$&$m_{eff}$/$m_0$&$\tau_q(ps)$&$\mu_q(cm^2V^{-1}s^{-1})$    &$\phi_B$&$k_F(\AA)$&$v_F(10^6ms^{-1})$&g-factor  \\ \hline
     $\alpha$&75.5&4.1&0.117&0.293&4386&1.92$\pi$&0.048&0.471&6.03   \\
     $\gamma$&510.4&4.5&0.167&0.269&2832&1.93$\pi$&0.124&0.864&1.13 \\
     \end{tabular}
     \begin{tabular}{cccccccccc}&
     \multicolumn{9}{c}{$\mathrm{BaGa_2}$}\\ 
     Index&$F(T)$&$T_D(K)$&$m_{eff}$/$m_0$&$\tau_q(ps)$&$\mu_q(cm^2V^{-1}s^{-1})$    &$\phi_B$&$k_F(\AA)$&$v_F(10^6ms^{-1})$&g-factor  \\ \hline
     $\alpha$&36.8&0.17&0.149&1.64&85830&1.92$\pi$&0.033&0.333&8.07   \\
     $\beta$&54.8&0.62&0.174&0.92&19734&0.97$\pi$&0.041&0.349&1.85\\
     $\gamma$&356.5&1.31&0.167&0.62&20661&1.95$\pi$&0.104&0.928&0.866\\
    \end{tabular}
    \caption{\label{table:1}Parameters extracted from dHvA oscillations of $\mathrm{SrGa_2}$ and $\mathrm{BaGa_2}$ for H $\parallel$ c for the fundamental oscillation frequencies. F is the oscillation frequency; T$_D$ is the Dingle temperature;  $\tau_q$ is the quantum lifetime; $\mu_q$ is the quantum mobility; $k_F$ and $v_F$ are the Fermi vector and Fermi velocity; $\phi_B$ is the Berry phase; g is the Landé g-factor.}
    \end{ruledtabular}
     
    \label{tab:my_label}
    
\end{table*}

The LK formula in Eq. \ref{eq1} is used to fit the QOs for both compounds with the inclusion of higher harmonics. We start with the fittings of the effective masses from the oscillatory amplitude for each pocket at different temperatures. All pockets in $\mathrm{SrGa_2}$ and $\mathrm{BaGa_2}$ have small effective masses, implying possible linear band crossings in these compounds. The Dingle temperature of each pocket could be obtained by fitting with these effective masses. We further add higher harmonics into our fitting, seeking to probe the phase terms (g, $\phi_B$, and $\delta$). Examples of the LK fitting on both compounds can be seen in Fig.~\ref{fig:5}(d,h). The good match between experimental data and LK fitting, as well as the small FFT amplitude of the residual signal (Fig.~\ref{fig:SI3}), both indicate good fitting quality. The $\delta$ term is a constant in each harmonic of the same frequency and can be uniquely determined. We intend to use the ratio of the oscillation amplitude and phases of higher harmonics to fit the Berry phase and g-factor. However, we find that the g and $\phi_B$ are not uniquely determined by these fits. More specifically, $\phi_B$ could differ by a $\pi$ between different combinations of g and $\phi_B$, as a consequence of the sinusoidal functions in the LK formula. This is illustrated by the following equality:
\begin{multline}
\label{eq2}
\cos(\psi_r) \sin(\phi_r)=
\cos(\pi r-\psi_r) \sin(\phi_r+\pi r)=\\ 
\cos(\pi r+\psi_r) \sin(\phi_r+\pi r)=
\cos(2\pi r-\psi_r) \sin(\phi_r)
\end{multline}
where $\psi_r={r\pi g\mu}/{2}$ and $\phi_r=2\pi \big[r({F}/{B}-{1}/{2}+\beta)+\delta\big]$ in Eq.~\ref{eq1}.
From Eq.~\ref{eq2}, it can be seen that for each ($g_0$, $\phi^0_B$) pair, there are three other pairs of ($g$, $\phi_B$) that yield the same LK fitting: ($2/\mu\pm g_0$, $\phi^0_B+\pi$), ($4/\mu- g_0$, $\phi^0_B$). We include one combination of g and $\phi_B$ in Table~\ref{table:1}. 

Therefore, we can only determine possible combinations of the Berry phase $\phi_B$ and the Landé g-factor rather than unique solutions, which hinders the exact determination of the Berry phase. To unambiguously refine the g and $\phi_B$, other approaches to measure the g and/or $\phi_B$ are required. 

In previous studies, peak splittings in the QO peaks or the FFT spectrum of the QOs have been attributed to the Zeeman effect. Such peak splittings are used to extract the Landé g-factor ~\cite{Nimori1994, Narayanan2015, Cao2015, HuJ2016, Liu2016, Martin2018, Xiang2021}. The splittings become more significant under larger magnetic fields (lower LLs). Surprisingly, no such splittings in the QO peaks were observed for the small Fermi pockets in the QOs of $\mathrm{SrGa_2}$ and $\mathrm{BaGa_2}$ even up to 65 T. For the $\gamma$ pocket in $\mathrm{SrGa_2}$ and $\mathrm{BaGa_2}$, we found more peaks within one period (Fig.~\ref{fig:SI5}) than previous reports~\cite{Narayanan2015, HuJ2016, Liu2016, Xiang2021}. These observations are inconsistent with the expectation of peak splittings. We also did not observe peak splittings in the FFT spectrum [Fig. \ref{fig:5}(b,f)]. Therefore, it is necessary to examine whether peak splittings are expected in $\mathrm{SrGa_2}$ and $\mathrm{BaGa_2}$ by revisiting the LK formula.

In materials that preserve {\it T} and {\it I}, the electronic bands are doubly degenerate. Under an external magnetic field, the Zeeman effect induces band-splitting between spin-up and spin-down bands, with the energy difference $\Delta E=2*gS\mu_B B$, where S is the electronic spin, $\mu_B$ is the Bohr magneton, and B is the external magnetic field. Such band splitting leads to a difference in the Fermi surface cross-section and has been treated as the underlying reason for peak splitting in QOs and their FFT spectra. The Landé g-factor can be estimated through the splitting~\cite{Narayanan2015, Liu2016, Xiang2021}. $\mathrm{SrGa_2}$ and $\mathrm{BaGa_2}$ satisfy the above description. However, we will demonstrate that these features might not be related to the Zeeman effect, since {\it T} and {\it I} are preserved in the compound. Instead, similar features may arise from other effects.

\subsection{LK formula for parabolic band dispersion}

We start from the LK formula, which is developed under the assumption of doubly-degenerate parabolic bands with energy dispersion $E_F={\hbar^2k^2_F}/{2m_{eff}}$. The following derivations are carried on QOs of magnetization, which could also be generalized to other physical properties. For parabolic bands, the corresponding frequency $F$ can be written as:
\begin{equation*} F=\frac{\hbar A}{2\pi e}=\frac{\hbar \pi k^2_F}{2\pi e}=\frac{m_{eff}E_F}{\hbar e} \end{equation*}
In the LK formula, $R_T$ and $R_D$ are amplitude damping terms that are consequences of Fermi surface broadening from finite temperature and LL broadening from finite quasiparticle lifetimes, respectively~\cite{Shoenberg1984}. The $R_S$ term is the result of the Zeeman effect, which will be determined in the following derivations. 

Under an external magnetic field, the degeneracy of the electronic bands is lifted $E^{\pm}=E_F\pm gS\mu_B B$, as the Zeeman energy scale ($\mu_B$ = 0.058 meV/T) is a few orders of magnitude smaller than the difference between the Fermi energy and the band top (which is usually on the order of 10-100 meV). Through the Onsager relationship, $F^{\pm}=F\pm {g\mu B}/{4}$. The overall QOs are the sum of contributions from spin-up and spin-down: 
\begin{multline}
\label{eq3}
\Delta M \propto \frac{1}{2}\big[\sin(\phi_r^+)+\sin(\phi_r^-)\big]= \\ \allowdisplaybreaks \cos\bigg(\frac{\phi_r^+-\phi_r^-}{2}\bigg)\sin\bigg(\frac{\phi_r^++\phi_r^-}{2}\bigg)=\cos(\psi_r)\sin(\phi_r)
\end{multline}
where $\phi_r^+=2\pi \big[r({F}/{B}-{1}/{2}+\beta)+\delta+ {gr\mu}/{4}\big]=\phi_r+\psi_r$ and $\phi_r^-=2\pi \big[r({F}/{B}-{1}/{2}+\beta)+\delta- {gr\mu}/{4}\big]=\phi_r-\psi_r$. Eq.~\ref{eq3} takes the exact form described by the LK formula~\cite{Shoenberg1984}. This assumes that the spin-up and spin-down contributions are equal, which might not hold true under strong fields where spin polarization becomes significant.

Next, we analyze a general case of different contributions from spin-up and spin-down electrons. The relative amplitude $\mathrm{{1}/{2}}$ is replaced by $\alpha^+$ and $\alpha^-$ with $\alpha^+ +\alpha^-=1$. The $\alpha^+$ and $\alpha^-$ are two general coefficients that could be field-dependent and reflect the relative concentrations of spin-up and spin-down electron concentrations. The expression in Eq.~\ref{eq3} is modified to:

\begin{multline}
\label{eq4}
\Delta M \propto \alpha^+\sin(\phi_r^+)+
\alpha^-\sin(\phi_r^-)
= C_r\sin(\phi_r+\epsilon)
\end{multline}
with 
\begin{equation*}
\label{eq5}
C_r=\sqrt{{\cos^2(\psi_r)}+(\alpha^+-\alpha^-)^2\sin^2(\psi_r)}\le1
\end{equation*}
 and 
 \begin{equation*}
 \label{eq6}
 \tan(\epsilon)=\frac{(\alpha^+-\alpha^-)\sin(\psi_r)}{\cos(\psi_r)}
\end{equation*}
The detailed derivation can be found in Appendix B. The form of the equation is the same as Eq.~\ref{eq3} after replacing the $R_S$ with $C_r$ and $\delta$ with $\epsilon+\delta$. This implies that the Zeeman effect does not change the overall QO frequency. The phase and absolute amplitude might change with the magnetic field since $\alpha^+$ and $\alpha^-$ could be field-dependent, but  $\alpha^+$ and $\alpha^-$ should only change gradually and should not register as QO peak splittings or frequency splittings. 

\subsection{LK formula for linear band dispersion}

In the previous analysis, we assumed parabolic electronic bands, where the quasiparticles are massive. We will move to the other limit: linear bands with massless quasiparticles.

The energy dispersion of a massless quasiparticle in a crystal lattice is $E=\hbar kv$, where $k$ and $v$ are the quasiparticle momentum and Fermi velocity, respectively. The QO frequency can be expressed as:
\begin{equation*} F=\frac{\hbar A}{2\pi e}=\frac{\hbar \pi k^2_F}{2\pi e}=\frac{{E_F}^2}{2\hbar e{v_F}^2} \end{equation*}  
Since the Zeeman effect is a few orders of magnitude smaller than the difference between the Fermi energy and the band top (which is usually on the order of 10-100 meV), it can also be treated as a weak perturbation: $E^{\pm}=E_F\pm gS\mu_B B$. The frequency can be written as:
\begin{equation}
    F^{\pm}=F\pm \frac{g\hbar k_F B}{4m_ev_F}+\frac{{(gB)}^2\hbar e}{32{(m_ev_F)}^2}
\end{equation}
The Zeeman effect introduces a term proportional to B similar to that of parabolic bands, while the effective mass ratio $\mu$ is replaced by a momentum ratio ${\hbar k_F}/{m_ev_F}$. This can be interpreted as the momentum of a quasiparticle scaled by the momentum of a free electron moving at the Fermi velocity. In contrast to the parabolic case, an additional $B^2$ term appears in the frequency for the linear case. Such a $B^2$ term is a negligible second-order term (the Zeeman effect and the linear in B term are first-order) for both spin-up and spin-down electrons. Therefore, $\Delta M$ for a linear band can be reproduced after substituting the following two terms in Eq.~\ref{eq1}: 
\begin{equation*}
    F\leftrightarrow F+\frac{{(gB)}^2\hbar e}{32{(m_ev_F)}^2}, \mu \leftrightarrow \frac{\hbar k_F}{m_ev_F}
\end{equation*}

The expression of dHvA oscillations $\Delta M$ in Eq.~\ref{eq3} and Eq.~\ref{eq4} becomes:
\begin{equation}
\label{eq7}
\Delta M \propto \cos(\psi'_r)\sin(\phi'_r)
\end{equation}
and
\begin{equation}
\label{eq8}
\Delta M \propto C'_r\sin(\phi'_r+\epsilon')
\end{equation} 
with
$\psi'_r={r\pi g \hbar k_F}/{2m_ev_F}$, $\phi'_r=2\pi \big[r({F}/{B}+{g^2B\hbar e}/{32(m_ev_F)^2}-{1}/{2}+\beta)+\delta\big]$, and 

\begin{equation*}
\label{eq9}
C'_r=\sqrt{{\cos^2(\psi'_r)}+(\alpha^+-\alpha^-)^2\sin^2(\psi'_r)}\le1
\end{equation*}

 and 
 \begin{equation*}
 \label{eq10}
 \tan(\epsilon')=\frac{(\alpha^+-\alpha^-)\sin(\psi'_r)}{\cos(\psi'_r)}
\end{equation*}
The detailed derivation can be found in Appendix B. Eq.~\ref{eq7} and Eq.~\ref{eq8} have the same form as Eq.~\ref{eq3} and Eq.~\ref{eq4}, implying that, as in parabolic bands, the Zeeman effect does not cause peak splittings in the QOs of linear bands. 
\begin{figure*}
\includegraphics[width=0.96\textwidth]{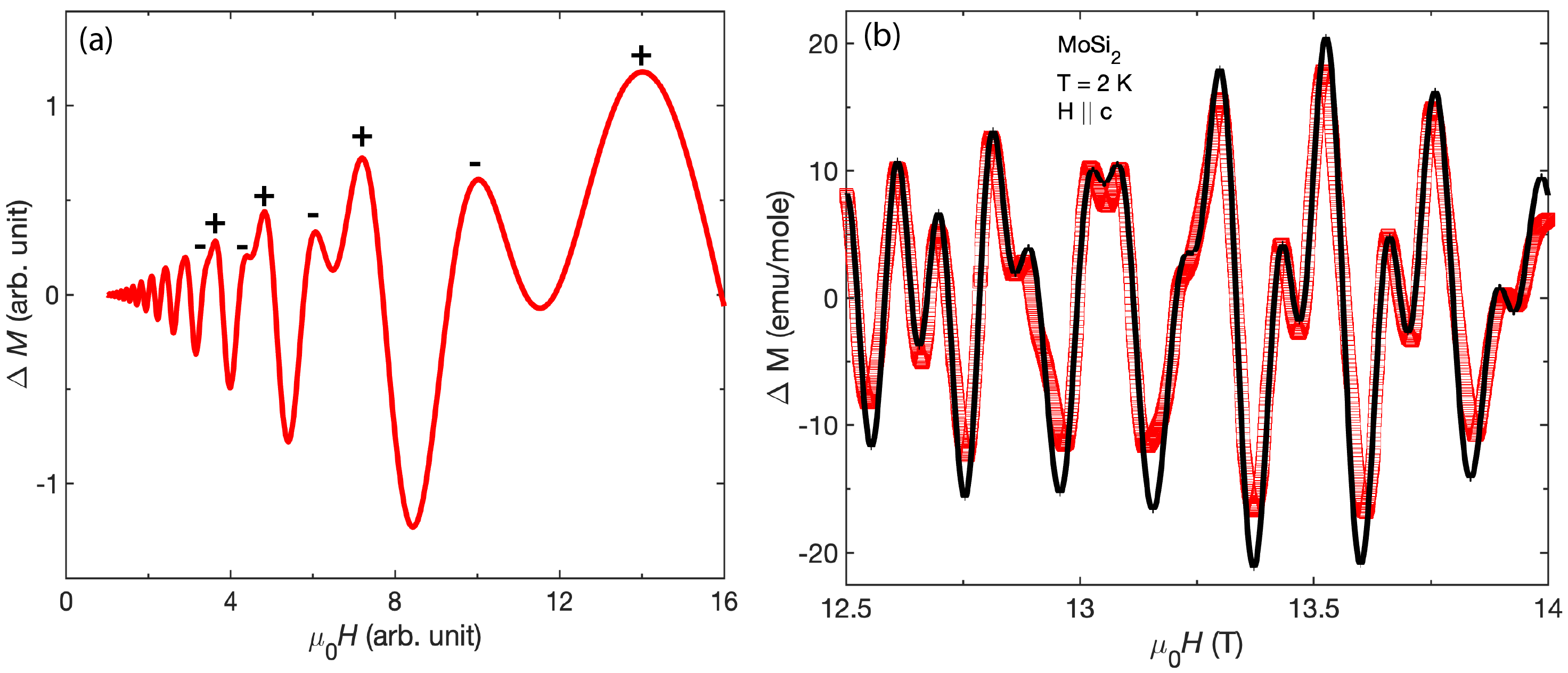} 
\caption{\label{fig:6} Simulations and fittings of dHvA oscillations. (a) Simulation of dHvA oscillations. The signal is simulated by fundamental oscillations of one frequency and the second harmonic. The +/- signs are signatures that were regarded as peak splitting. (b) dHvA oscillations of $\mathrm{MoSi_2}$ (red symbols) and LK fittings with higher harmonics (black line). The data is extracted from~\cite{Martin2018}. }
\end{figure*}
\subsection{Possible explanations for peak splitting in quantum oscillations}
Our analysis excludes the Zeeman effect as the underlying reason for peak splitting in compounds that preserve {\it T} and {\it I}. We provide two alternative explanations for the peak splitting in QOs: 1. QO peak splitting and 2. peak splitting in FFT~\cite{Nimori1994, Narayanan2015, Cao2015, HuJ2016, Liu2016, Martin2018, Xiang2021}.

The QO peak splitting is manifested as the two-peak feature at low LLs (at large magnetic fields), which has been observed and reported in a few Dirac semimetals, including $\mathrm{Cd_3As_2}$, $\mathrm{ZrTe_5}$, $\mathrm{ZrSiS}$, $\mathrm{MoSi_2}$ ~\cite{Narayanan2015, Cao2015, Liu2016, Hu2017,Martin2018, Xiang2021}. The difference between the inverse field $\Delta({1}/{B})={1}/{B^+}-{1}/{B^-}$ of the two peak positions is found to be a constant, directly related to the Landé g-factor~\cite{Narayanan2015}. 

As the Zeeman effect becomes more pronounced at large magnetic fields, the amplitude of the second harmonic increases dramatically. The frequency and phase of the second harmonic are different from those of the fundamental oscillation: the frequency is doubled, and there is a phase change $\Delta \Phi=\pi-\beta$ or $2 \pi-\beta$, depending on the sign of the spin-damping terms in both the fundamental and the second harmonic oscillations. The sum of these two signals could lead to the two-peak feature which is identical to the peak splitting. This peak splitting can be seen in the simulation shown in Fig.~\ref{fig:6}(a). 

For reference, we carried out an LK-fitting on measured dHvA oscillations in $\mathrm{MoSi_2}$ (Fig.~\ref{fig:6}(b)), where peak splitting was reported~\cite{Martin2018}. The fitting matches all the features attributed to peak splittings with fundamental oscillations and their second harmonics of the two frequencies (Fig.~\ref{fig:6}(b)). Moreover, the peak splitting in $\mathrm{MoSi_2}$ is only reported at temperatures where the FFT of the second harmonics does not vanish~\cite{Martin2018}. Hence, what is observed might be the superposition of fundamental oscillations and their second harmonics rather than the Zeeman effect.

Peak splittings in FFT appear as two adjacent peaks with similar amplitude in the FFT spectrum. In addition, the difference between the peak positions is often found to change with the range of the FFT window, which seems plausible as the Zeeman energy increases linearly with the magnetic field \cite{Nimori1994}. However, based on previous derivations, the Zeeman effect should not induce frequency changes in the QO signals. We point out that other intrinsic factors might generate peak splittings in FFT. The two peaks might correspond to oscillations from two distinct Fermi surface cross-sections. If the band characters (band curvature, quasiparticle effective mass, quantum mobility) are similar, the amplitude of these two oscillations will be close. That has been the case in $\mathrm{MoSi_2}$, where the $\beta_1$ and $\beta_2$ peaks are proposed to be the minimum and maximum cross-sections of a quasi-2D Fermi pocket, respectively~\cite{Pavlosiuk2022}. In general, these could also correspond to two distinct Fermi pockets.

The two peaks in the FFT spectrum might also be split by magnetic interaction, with or without magnetic order. In materials with magnetic ions, an external magnetic field or the magnetic order breaks {\it T} and lifts the degeneracy of the energy bands. The exchange interaction between the magnetic moments induces energy shifts to the conduction electron bands, which are registered as splittings in the QO frequencies. In contrast to the Zeeman effect, the exchange interaction is generally much stronger. The energy shifts from magnetic interactions are not proportional to the external field, which may lead to peak splittings in the FFT spectrum. Such a scenario may apply to TmSb, which is a centrosymmetric paramagnetic compound down to 0.5 K \cite{Vogt1968, Nimori1995}. Based on the isothermal magnetization measurements, the Tm ions carry a local moment and the magnetization does not saturate up to 10 T \cite{Vogt1968, Nimori1995}. In addition, conduction electrons mediate the RKKY interaction between $\mathrm{Tm^{3+}}$ ions in TmSb. These findings suggest that the magnetic interaction shifts the conduction electron bands, which could account for the different frequencies in the FFT spectrum of different FFT windows observed in~\cite{Nimori1994}. 

\section{Discussion}
Based on the self-consistent study of the Fermi surface and band structure through DFT, ARPES, and angle-dependent QOs, we have established a comprehensive understanding of the Fermi surface topology on $\mathrm{BaGa_2}$ and $\mathrm{SrGa_2}$. The Fermi pocket centered at the K-point of the Brillouin zone is related to the Dirac nodal line, which is the only topologically nontrivial band crossing in these materials that intersects the Fermi level. Furthermore, the large non-saturating magnetoresistance, non-compensated carrier nature, and PHE are observed in these materials. These transport properties are associated with the topological band crossings in $\mathrm{BaGa_2}$ and $\mathrm{SrGa_2}$ \cite{Hu2019}. 

We further demonstrate that the accurate and rigorous determination of the Berry phase through QOs is challenging. The values of $\phi_B$ and g in Table~\ref{table:1} are chosen such that only the $\phi_B$ for the $\beta$ frequency in $\mathrm{BaGa_2}$ is close to $\pi$, as it is the only topologically nontrivial Fermi pocket. Note that the Landé g-factor can be estimated through calculations \cite{ZhangS2023}, and may serve as a guide to disentangle $\phi_B$ and g.

The spin-zero effect has been shown to be another method to estimate the Landé g-factor~\cite{Wang2018, Bi2018}. It refers to the scenario where the spin damage terms satisfy $R_S^{1st}=\cos({\pi g\mu}/{2})=0$ and $R_S^{2nd}=\cos({2\pi g\mu}/{2})=\pm1$. If this condition is met, the fundamental oscillations vanish, while the second harmonic oscillations do not, and thus provide a way to calculate the Landé g-factor. However, the spin-zero effect has only been reported in a few compounds at certain magnetic field orientations~\cite{Wang2018, Bi2018}, and is absent in most quantum materials. In $\mathrm{SrGa_2}$ and $\mathrm{BaGa_2}$, we did not observe any spin-zero effect. Moreover, the Landé g-factor and effective mass are anisotropic. The g-factor obtained at the spin-zero orientation cannot be generalized to be utilized at other orientations. However, this may serve as an estimate of the value of the overall g-factor.

\section{Conclusions}
In summary, our results demonstrate that $\mathrm{BaGa_2}$ and $\mathrm{SrGa_2}$ are Dirac nodal line semimetals through a comprehensive Fermi surface study via a combination of magnetotransport, ARPES, and DFT calculations. By implementing an LK fitting including higher harmonics, we found that the Berry phase $\phi_B$ and Landé g-factor are entangled and cannot be uniquely determined. Through further elaboration on the LK formula and data simulation, we speculate that the Zeeman effect might not be the real reason for peak splitting in QOs. We highlight the challenges in the unambiguous determination of the Berry phase $\phi_B$ in QOs, which might be overcome with a good estimate through calculations or the spin-zero effect. 

\begin{acknowledgments}
We thank Amalia Coldea for useful discussions. This work was primarily supported by the Department of Defense, Air Force Office of Scientific Research under Grant No. FA9550-21-1-0343.
EM and SL acknowledge partial support by the Robert A. Welch Foundation Grant No. C-2114. MY acknowledges partial support by the Gordon and Betty Moore Foundation's EPiQS Initiative through grant No. GBMF9470 and the Robert A. Welch Foundation Grant No. C-2175.
This research used resources of the Advanced Light Source, which is a DOE Office of Science User Facility under contract No. DE-AC02-05CH11231. Part of the research described in this paper was performed at the Canadian Light Source, a national research facility of the University of Saskatchewan, which is supported by the Canada Foundation for Innovation (CFI), the Natural Sciences and Engineering Research Council (NSERC), the National Research Council (NRC), the Canadian Institutes of Health Research (CIHR), the Government of Saskatchewan, and the University of Saskatchewan. The National High Magnetic Field Laboratory Pulsed-Field Facility is funded by the National Science Foundation Cooperative Agreement Number DMR-2128556, the State of Florida, and the U.S. Department of Energy. Measurements in pulsed fields were supported by the DOE-BES Science of 100 T program.
\end{acknowledgments}

\pagebreak

\newpage

\begin{appendices}

\setcounter{figure}{0}

\section{LK fitting and Berry phase analysis in $\mathrm{SrGa_2}$ with bandpass filter}

\begin{figure}
\includegraphics[width=0.48\textwidth]{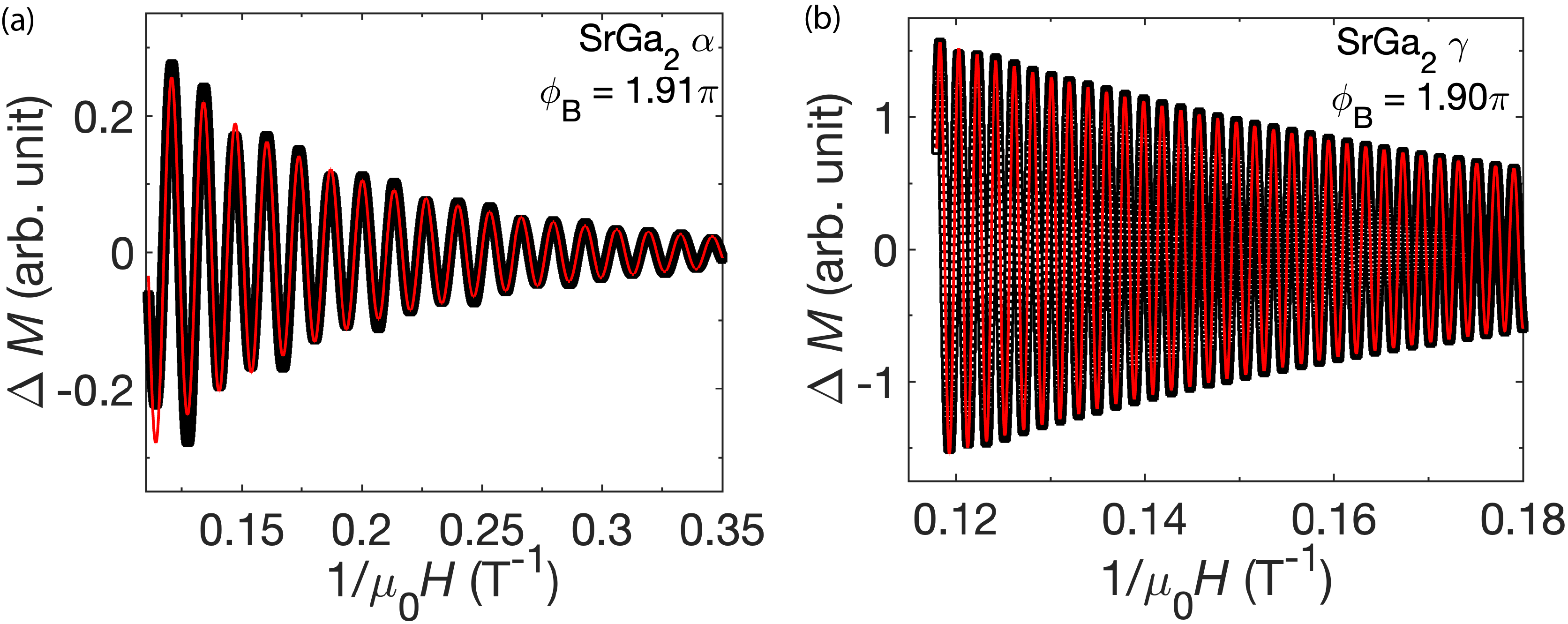}
\renewcommand{\thefigure}{S\arabic{figure}}
\caption{\label{fig:SIBP} LK fitting to QOs in $\mathrm{SrGa_2}$ after band-pass filter. (a) QOs with band-pass applied to the $\alpha$ frequency at 76 T (black) and the LK fitting (red). (b) QOs with band-pass applied to the $\gamma$ frequency at 510 T (black) and the LK fitting (red). }
\end{figure}

To further validate our direct LK fitting approach with the inclusion of higher harmonics, we performed analysis on signals after applying a band-pass filter to single out fundamental oscillations of $\alpha$ and $\beta$ frequencies with the magnetic field along c-axis, respectively. As band-pass filters tend to distort the oscillations at the boundary, the subsequent LK fitting is conducted in the intermediate field range. As shown in Fig. \ref{fig:SIBP}, a good fitting quality is achieved for both oscillations. The extracted Berry phases from the band-pass filtered results are almost the same as our results shown in the main text.

\section{Detailed derivations of LK formula for parabolic and linear band dispersion}


\begin{figure*}
\includegraphics[width=0.96\textwidth]{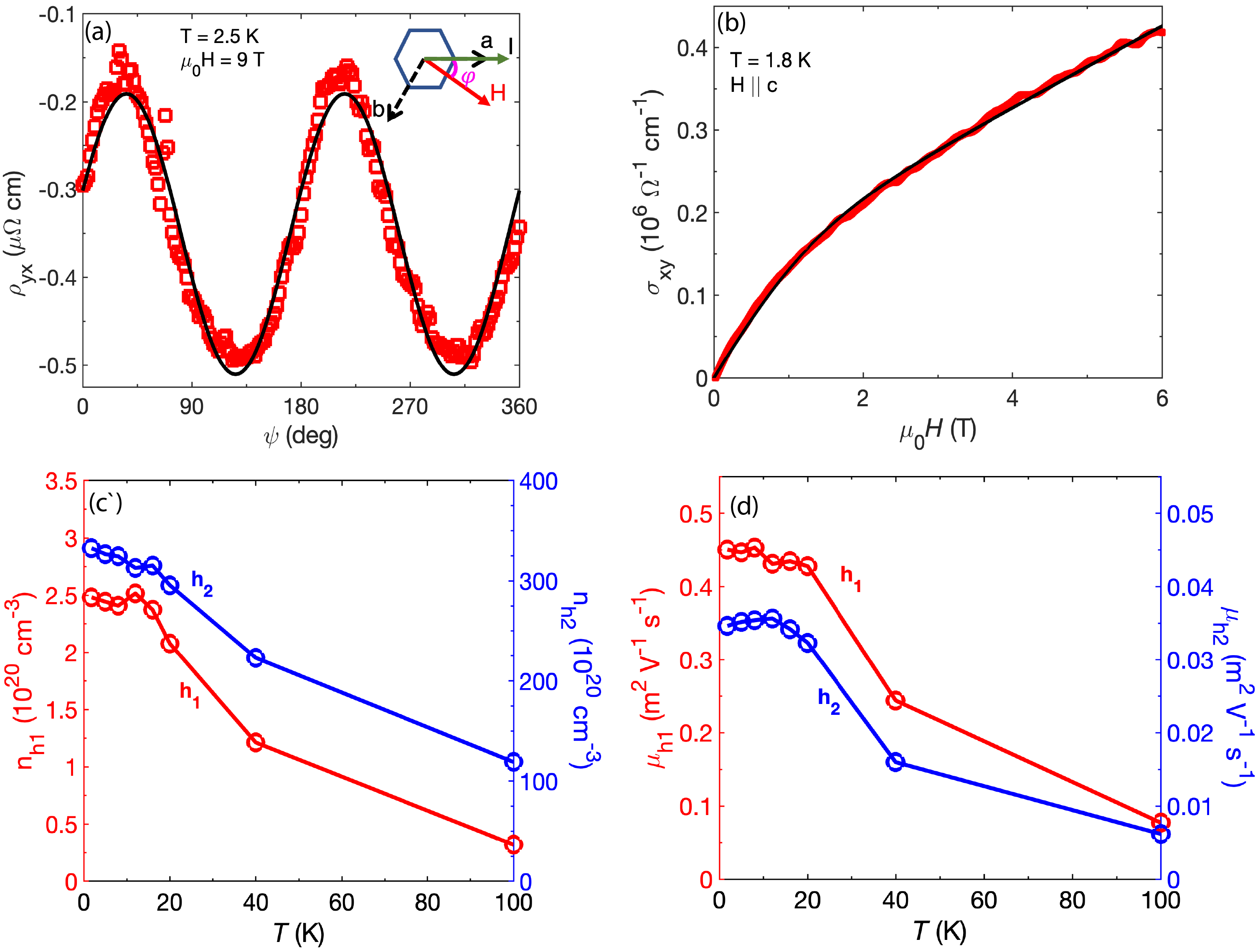}
\renewcommand{\thefigure}{S\arabic{figure}}
\caption{\label{fig:SI1} Hall conductivity under in-plane and out-of-plane magnetic fields of $\mathrm{SrGa_2}$. (a) Hall resistivity under in-plane magnetic field (red symbols) and fit of the planar Hall effect (black line) at 2.5~K, 9~T. The measurement geometry is shown in the inset.  (b) Hall conductivity under out-of-plane magnetic field (red symbols) at 1.8~K fitted by a two-band model with two hole carriers of different mobility (black line).  (c,d) Carrier concentration (c) and mobility (d) from the two-band model fitting of Hall conductivity under out-of-plane magnetic field. }
\end{figure*}

For the case of parabolic band dispersion with $\alpha^++\alpha^-=1$, the LK formula for magnetization is modified to:
\begin{multline}
\label{eq11}
\Delta M \propto \alpha^+\sin(\phi_r^+)+
\alpha^-\sin(\phi_r^-) \\ \allowdisplaybreaks
= \frac{1}{2}\big[\sin(\phi_r^+)+\sin(\phi_r^-)\big]+  \frac{\alpha^+-\alpha^-}{2}\big[\sin(\phi_r^+)-\sin(\phi_r^-)\big] \\ \allowdisplaybreaks = \sin\bigg(\frac{\phi_r^++\phi_r^-}{2}\bigg)\cos\bigg(\frac{\phi_r^+-\phi_r^-}{2}\bigg)  
+\\ \allowdisplaybreaks(\alpha^+-\alpha^-)\cos\bigg(\frac{\phi_r^++\phi_r^-}{2}\bigg)\sin\bigg(\frac{\phi_r^+-\phi_r^-}{2}\bigg) \\ \allowdisplaybreaks 
=\sin(\phi_r)\cos(\psi_r)+(\alpha^+-\alpha^-)\cos(\phi_r)\sin(\psi_r) 
\end{multline}
Let \begin{equation*}
\label{eq12}
C_r=\sqrt{{\cos^2(\psi_r)}+(\alpha^+-\alpha^-)^2\sin^2(\psi_r)}\le1
\end{equation*}
 and 
 \begin{equation*}
 \label{eq13}
 \tan(\epsilon)=\frac{(\alpha^+-\alpha^-)\sin(\psi_r)}{\cos(\psi_r)}
\end{equation*} Eq.~\ref{eq11} can be written as:
\begin{multline}
\label{eq14}
\Delta M \propto C_r\big[\sin(\phi_r)\cos(\epsilon)+\cos(\phi_r)\sin(\epsilon)\big] \\ \allowdisplaybreaks = C_r\sin(\phi_r+\epsilon)
\end{multline} which comes to same equation as Eq.~\ref{eq4}.

For the case of linear dispersion without consideration of spin polarization, the LK formula for magnetization is modified to: 
\begin{multline}
\label{eq15}
\Delta M \propto \frac{1}{2}\big[\sin({\phi '}^{+}_r)+\sin({\phi '}^{-}_r)\big]= \\ \allowdisplaybreaks
\cos\bigg(\frac{{\phi '}^{+}_r-{\phi '}^{-}_r}{2}\bigg)\sin\bigg(\frac{{\phi '}^{+}_r+{\phi '}^{-}_r}{2}\bigg)=\cos(\psi'_r)\sin(\phi'_r)
\end{multline}
where 
\begin{equation*}
{\phi '}^{+}_r=2\pi \bigg[r\bigg(\frac{F}{B}+\frac{g^2B\hbar e}{32(m_ev_F)^2}-\frac{1}{2}+\beta\bigg)+\delta+ \frac{gr\hbar k_F}{4m_ev_F}\bigg]={\phi'}_r+{\psi'}_r
\end{equation*}
and 
\begin{equation*}{\phi '}^{-}_r=2\pi \bigg[r\bigg(\frac{F}{B}+\frac{g^2B\hbar e}{32(m_ev_F)^2}-\frac{1}{2}+\beta\bigg)+\delta- \frac{gr\hbar k_F}{4m_ev_F}\bigg]={\phi'}_r-{\psi'}_r\end{equation*}

For the case with consideration of spin polarization with $\alpha^++\alpha^-=1$, the LK formula for magnetization is modified to:
\begin{multline}
\label{eq16}
\Delta M \propto \alpha^+\sin({\phi '}^{+}_r)+\alpha^-\sin({\phi '}^{-}_r)= \\ \allowdisplaybreaks
\frac{1}{2}\big[\sin({\phi '}^{+}_r)+\sin({\phi '}^{-}_r)\big]+  \frac{\alpha^+-\alpha^-}{2}\big[\sin({\phi '}^{+}_r)-\sin({\phi '}^{-}_r)\big] \\ \allowdisplaybreaks
=\sin\bigg(\frac{{\phi '}^{+}_r+{\phi '}^{-}_r}{2}\bigg)\cos\bigg(\frac{{\phi '}^{+}_r-{\phi '}^{-}_r}{2}\bigg) 
+\\ \allowdisplaybreaks(\alpha^+-\alpha^-)\cos\bigg(\frac{{\phi '}^{+}_r+{\phi '}^{-}_r}{2}\bigg)\sin\bigg(\frac{{\phi '}^{+}_r-{\phi '}^{-}_r}{2}\bigg) \\ \allowdisplaybreaks 
=\sin(\phi'_r)\cos(\psi'_r)+(\alpha^+-\alpha^-)\cos(\phi'_r)\sin(\psi'_r) 
\end{multline}

Let \begin{equation*}
\label{eq17}
C'_r=\sqrt{{\cos^2(\psi'_r)}+(\alpha^+-\alpha^-)^2\sin^2(\psi'_r)}\le1
\end{equation*}
 and 
 \begin{equation*}
 \label{eq19}
 \tan(\epsilon')=\frac{(\alpha^+-\alpha^-)\sin(\psi'_r)}{\cos(\psi'_r)}
\end{equation*} Eq.~\ref{eq16} can be written as:
\begin{multline}
\label{eq20}
\Delta M \propto C'_r\big[\sin(\phi'_r)\cos(\epsilon')+\cos(\phi'_r)\sin(\epsilon')\big] \\ \allowdisplaybreaks = C'_r\sin(\phi'_r+\epsilon')
\end{multline} which comes to same equation as Eq.~\ref{eq8}.

\begin{figure*}
\includegraphics[width=0.96\textwidth]{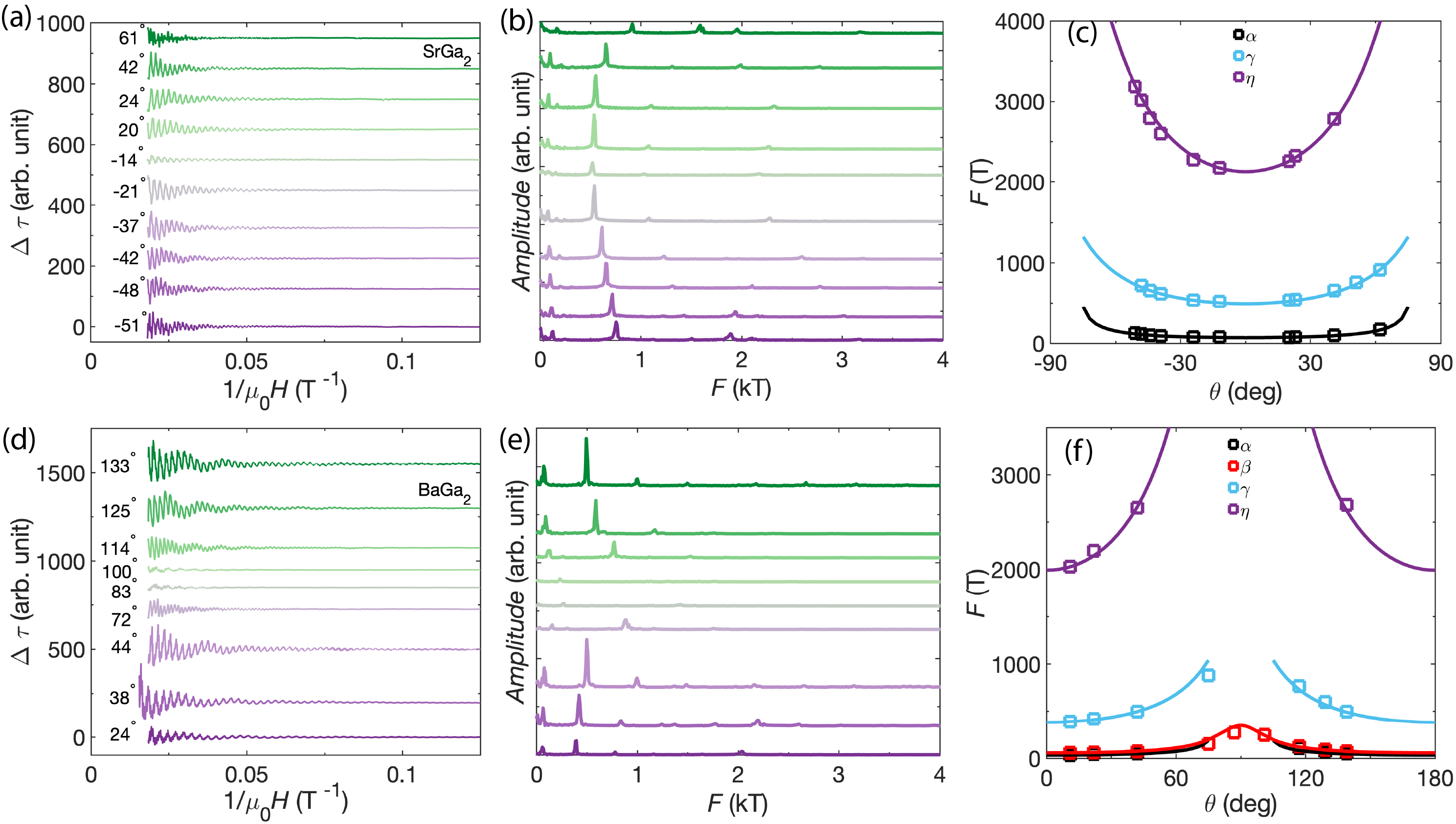}
\renewcommand{\thefigure}{S\arabic{figure}}
\caption{\label{fig:SI4} Pulse field angle-dependent magnetic torque measurements of $\mathrm{SrGa_2}$ and $\mathrm{BaGa_2}$. (a,d) Magnetic torque $\Delta \tau$ after smooth background subtraction of $\mathrm{SrGa_2}$ (a) and $\mathrm{BaGa_2}$ (d). The angle $\theta$ is defined in the same way as in Fig.~\ref{fig:3}. (b,e) FFT spectra of $\Delta \tau$ of $\mathrm{SrGa_2}$ (b) and $\mathrm{BaGa_2}$ (e). No signatures of peak splitting are observed in either background-subtracted data or the FFT spectra. (c,f) Fundamental QO frequencies as a function of angle for $\mathrm{SrGa_2}$ (c) and $\mathrm{BaGa_2}$ (f). The symbols are values extracted from FFTs in (b, e), the lines are values from DFT. }
\end{figure*}

\begin{figure*}
\includegraphics[width=0.96\textwidth]{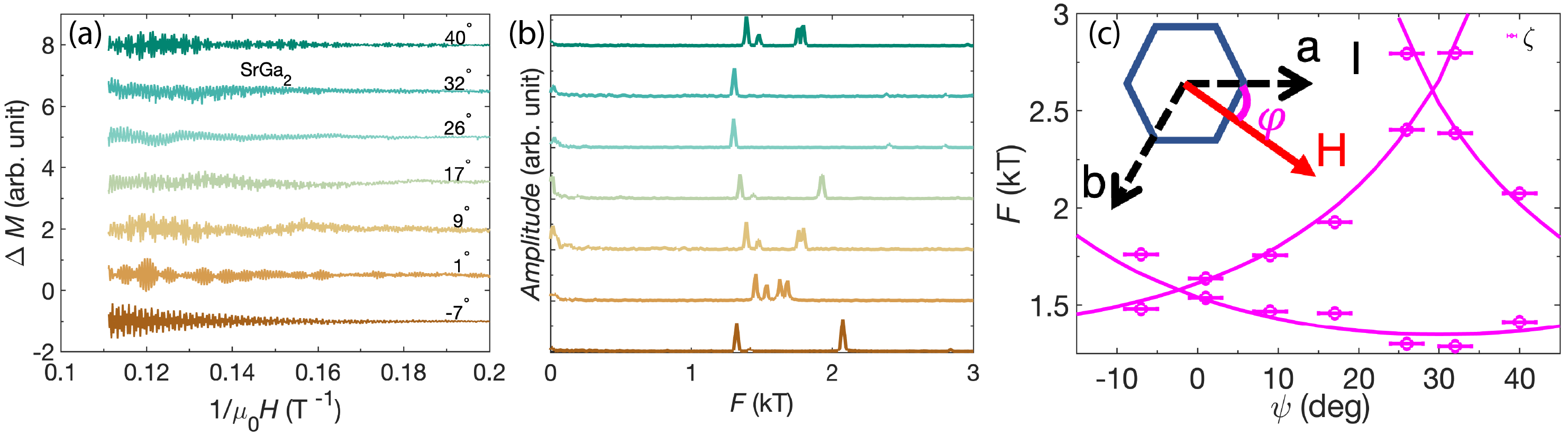}
\renewcommand{\thefigure}{S\arabic{figure}}
\caption{\label{fig:SI2} In-plane magnetic field angle dependent dHvA oscillations of $\mathrm{SrGa_2}$. (a) Angle-dependent dHvA oscillations of $\mathrm{SrGa_2}$ with different field orientations at 2.5 K. (b) FFT spectra of $\Delta M$ of $\mathrm{SrGa_2}$. (c) Frequencies from FFT (black symbols) and from calculations (solid lines) as a function of angle. The inset is a schematic of the measurement geometry. }
\end{figure*}

\begin{figure*}
\includegraphics[width=0.96\textwidth]{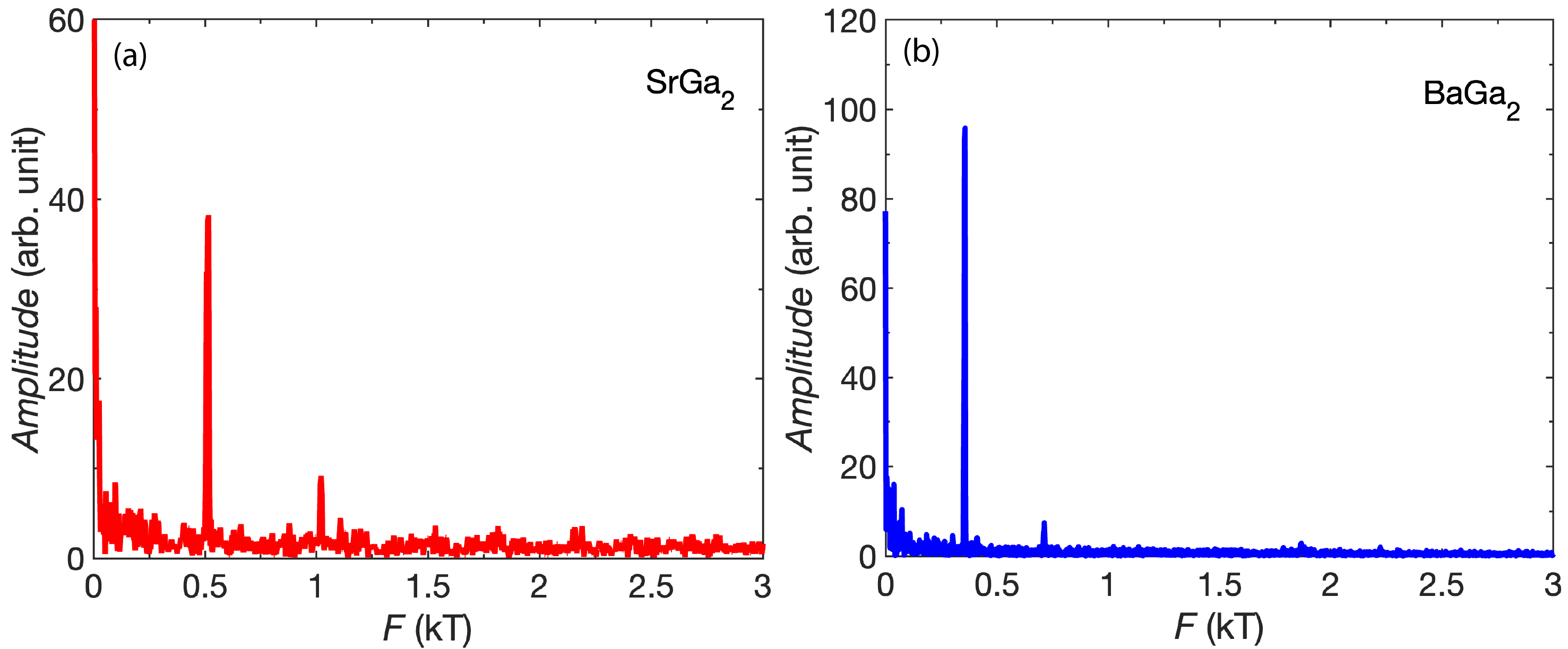}
\renewcommand{\thefigure}{S\arabic{figure}}
\caption{\label{fig:SI3} FFT of the residue (the difference between raw data and our LK fitting) of (a)$\mathrm{SrGa_2}$ and (b)$\mathrm{BaGa_2}$ after subtraction of the LK fitting at 2.5~K. The FFT amplitude in the residue is much smaller ($<$~3 \%) than the FFT of the original signal, indicating a very small difference between raw data and our LK fitting, further implying a very good fitting quality. 
}
\end{figure*}

\begin{figure*}
\includegraphics[width=0.96\textwidth]{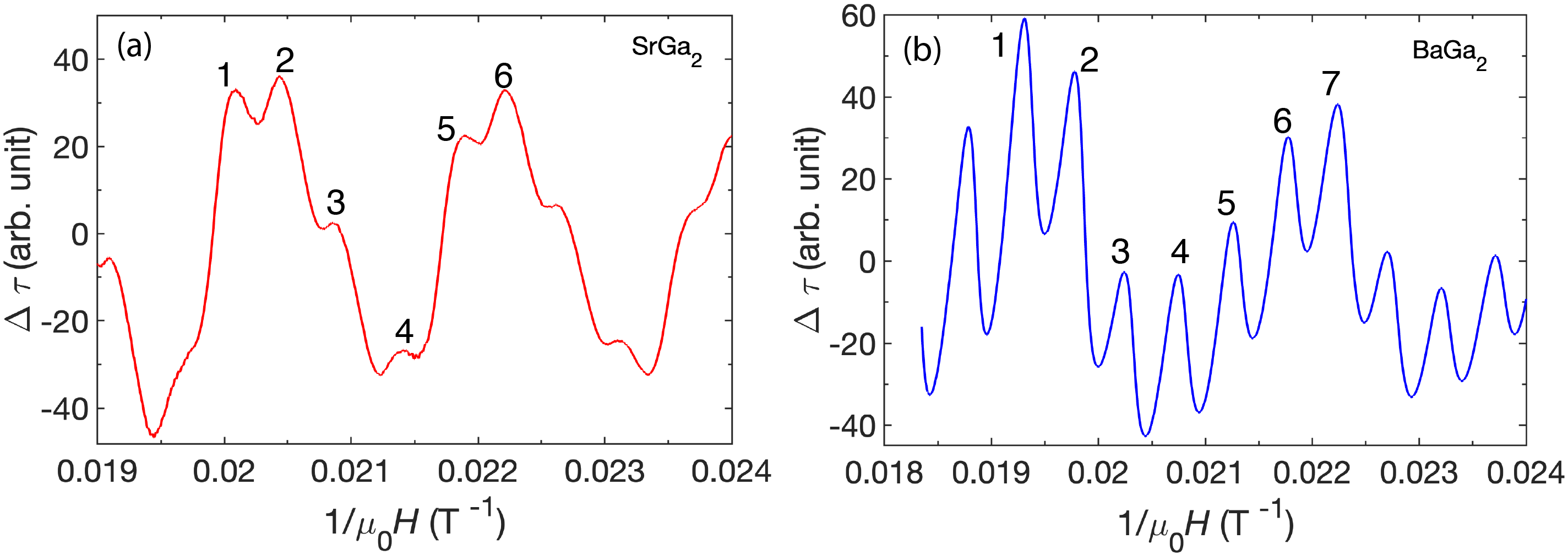}
\renewcommand{\thefigure}{S\arabic{figure}}
\caption{\label{fig:SI5} High field segments of magnetic torque on $\mathrm{SrGa_2}$ (a) and $\mathrm{BaGa_2}$ (b). The peaks within one period of the quantum oscillations of Fermi pocket $\gamma$ in $\mathrm{SrGa_2}$ and $\gamma$ in $\mathrm{BaGa_2}$ are labeled with numbers.}
\end{figure*}

\end{appendices}
\end{document}